\documentclass{article} 
\usepackage{amsmath}
\usepackage{amsfonts,amssymb}
\usepackage{graphicx}
\usepackage{epsfig}

\newcommand{\D}{\displaystyle}

\newcommand{\sgn}{{\rm sgn}}
\newcommand{\tr}{{\rm tr}\,}
\newtheorem{theorem}{Theorem}
\newtheorem{proposition}{Proposition}

\title{Exact diagonalisation of 1-d interacting spinless Fermions}

\author{Heiner Kohler\thanks{heinerich.kohler@uni-due.de}\\
\small{ Fakult{\"a}t f{\"u}r Physik,  Universit\"at Duisburg--Essen,}\\
\small{ Lotharstrasse 1-21, 47057 Duisburg, Germany.}
} 
\begin{document}

\maketitle
\begin{abstract}
We acquire a method of constructing an infinite set of exact eigenfunctions of 1--d interacting spinless Fermionic systems. Creation and annihilation operators for the interacting system are found and thereby the many--body Hamiltonian is diagonalised. The formalism is applied to several examples. One example is the theory of Jack polynomials. For the Calogero-Moser-Sutherland Hamiltonian a direct proof is given that the asymptotic Bethe Ansatz is correct.
\end{abstract}

\section{Introduction}
\label{sec1}
The study of one--dimensional integrable models of interacting particles has a long history going 
back to Bethe \cite{bet31}. In physics there has been a renewed interest in one--dimensional integrable systems recently in the study of cold atom gases and Bose--Einstein condensates \cite{fuc04,che04a,che04b}. 
The standard way of constructing eigenfunctions is Bethe's Ansatz. One crucial condition for Bethe's Ansatz to be successful is that the two--body scattering matrix $S_{ab}(k_j-k_k)$ of a particle of type $a$ with momentum $k_j$ with a particle of type $b$ with momentum $k_k$ fulfills the 
Yang--Baxter (star triangle) equation. The Yang--Baxter equation serves as the starting point for the algebraic Bethe Ansatz. For details about the Yang--Baxter equation and on the algebraic Bethe Ansatz see for instance \cite{bax89,kor93} and references therein.

Imposing periodic boundary conditions for the $N$ particle wave function, from Bethe's Ansatz one (or a set) of Fredholm integral equations for the density of states is obtained. They are referred to as {\em Bethe Ansatz equations}. 
Applied either to 1--d quantum mechanical, or to 2--d classical lattice theories, 
Bethe's Ansatz has been extraordinarily successful. For one reason, because in many of the most important lattice models, as for instance the Hubbard model, Heisenberg model, Ising model etc. only nearest neighbor interaction is assumed. Applied to continuous models Bethe's Ansatz is in its simplest form constrained to particles with $\delta$--interaction \cite{lie63,yan67,gau66}, being the only strictly local interaction. 

The problems, connected with non--local interactions in continuum models, were partly overcome by the asymptotic Bethe Ansatz (ABA). It was introduced by Sutherland \cite{sut71,sut71b} in order to obtain thermodynamical quantities for  Calogero--Moser--Sutherland (CMS) models. 

The basic assumption of ABA is that the Bethe Ansatz equations still hold for non--local interactions as long as the $N$--body $S$--matrix factorizes into a product of $2$--body $S$--matrices. Using the ABA hypothesis it is therefore possible to obtain thermodynamical quantities without detailed knowledge of the wave function in the interacting region. This assumption has been proven to be correct for the trigonometric CMS model, where a complete set of eigenfunctions can be constructed with Jack polynomials \cite{mac92,for92,for93,for95}. In that specific model even some thermodynamical correlation functions could be calculated \cite{ha95}. 

Although the ABA has been proven to be correct by other methods \cite{sut95} and in some other cases \cite{kaw92}, the different treatment of local and non--local interactions within the Bethe Ansatz is unsatisfactory. Moreover the ABA yields no clue of how to construct eigenfunctions of systems with non--local interaction. Therefore in this work I wish to put forward an approach, alternative  to Bethe's Ansatz, which treats local and non--local interactions on the same basis. 

The exact $N$--body wave function is not constructed via an Ansatz and by adjusting parameters but by the successive application of a creation operator onto the vacuum ground state. Thereby an integral representation for an arbitrary $N$ particle eigenfunction is obtained. The explicit construction of this creation operator and its corresponding annihilation operator for certain interaction potentials  is my main result.

Since the eigenfunctions are constructed in coordinate basis the method shares features with the original coordinate Bethe Ansatz \cite{bet31,lie63}. On the other hand the construction resembles the algebraic Bethe Ansatz \cite{kor93} inasmuch a vacuum state is successively filled by the action of a creation operator. Indeed, bridging the gap between coordinate Bethe Ansatz and algebraic Bethe Ansatz was one motivation for this work.

The basis idea is rooted in the following observation: a class of multidimensional integration formulae -- some of them are known for a long time -- permits a natural interpretation as Fermionic creation operators for a one--dimensional many--body Hamiltonian. These integrals share the common property that the integration domain $I_{\rm in}^{(N)}$ of a set ${\bf x}^\prime$ of $N$ integration variables $x_n^\prime$, $1\leq n\leq N$ is defined by the interlacing condition
\begin{equation}
\label{interlacing}
x_1> x_1^\prime > x_2 > \ldots>x_N> x_N^\prime> x_{N+1} \ .
\end{equation}
The integral itself is therefore in general a function of a set ${\bf x}$ of $N+1$ arguments $x_i$. Due to condition (\ref{interlacing}) this function vanishes, whenever two arguments are equal. By the same token it is antisymmetric under a permutation of arguments, if the integrand is antisymmetric in the primed and symmetric in the unprimed variables. Thus by 
construction the integral has the nature of a Fermionic wave function, when the set ${\bf x}$ is interpreted as the particle positions. 

To illustrate this idea let us consider the following version of the Dixon--Anderson integral \cite{dix05,and91} 
\begin{eqnarray}
\label{anderson}
\frac{\Gamma^{N+1}(\lambda+1)} {\Gamma((N+1)(\lambda+1))} & = &
                             \int_{I_{\rm in}^{(N)}} \prod_{i=1}^N dx_i^\prime \mu_\lambda({\bf x},{\bf x}^\prime) \\
\label{anderson1}\mu_\lambda({\bf x},{\bf x}^\prime) &=& \frac{\Delta_N({\bf x}^\prime)}{\Delta_{N+1}^{2\lambda+1}({\bf x})}  
                                \prod_{i=1}^{N}\prod_{j=i}^N(x_i-x_j^\prime)^{\lambda}
                                \prod_{j=1}^{N}\prod_{i=j+1}^{N+1}(x_j^\prime-x_i)^{\lambda} 
                                  \ .
\end{eqnarray}
If both sides of the above equation are multiplied with $\Delta_{N+1}^{2\lambda+1}({\bf x})$, it is a representation of a power of Vandermonde's determinant 
\begin{equation}
\Delta_{N+1}({\bf x}) \equiv \prod_{n<m}^{N+1}(x_n-x_m)
\end{equation}
as an integral over $\Delta_{N}({\bf x}^\prime)$. 

As a second example let us follow Ref.~\cite{guh02a} and consider group integrals of the form 
\begin{equation} 
\phi^{(\lambda)}_N({\bf k},{\bf x})\ = \ \int_{U\in U(N)} d\mu(U)\exp(-i{\rm tr}U^{-1}{\bf x}U{\bf k}) \ .
\end{equation}
Here $U(N)$ is a classical compact Lie group manifold and ${\bf x}$, ${\bf k}$ are diagonal $N\times N$ matrices. 
The Haar measure of the group is denoted by $d\mu(U)$. The parameter $\lambda$ depends on the group under consideration\footnote{\label{fn1}$\lambda$ is related to the parameter $\beta$ of \cite{guh02a} by $\lambda=\beta/2-1$.}. 
It was found that $\phi^{(\lambda)}_N({\bf k},{\bf x})$ can be constructed recursively in the dimension $N$ of the group as
\begin{equation}
\label{recurs}
\phi^{(\lambda)}_{N+1}({\bf k}^+,{\bf x}) \ \propto \ \int_{I_{\rm in}^{(N)}}\prod_{i=1}^N dx_i^\prime \mu_\lambda({\bf x},{\bf x}^\prime) e^{ik_{N+1} (\tr x-\tr x^\prime)}
            \phi^{(\lambda)}_{N}({\bf k},{\bf x}^\prime) \ ,
\end{equation}
where $k_{N+1}$ is the additional element of the new $(N+1)\times (N+1)$ matrix ${\bf k}^+$ on the left hand side.

As a third example I mention the integration formula for symmetric Jack polynomials found by Okounkov and Olshanski \cite{oko97} and by Kuzentsov {\em et al.} \cite{kuz03}. Let 
$J^{(\lambda)}_{N}({\bf n},{\bf x})$ be a symmetric Jack polynomial in $N$ variables\footnote{\label{allmbdarel} $\lambda$ is related to the parameter $\alpha$ of MacDonalds book by $\lambda=1/\alpha-1$.} then 
\begin{equation}
\label{jack}
J^{(\lambda)}_{N+1}({\bf n},{\bf x}) \ \propto \ 
     \int_{I_{\rm in}^{(N)}} \prod_{i=1}^N dx_i^\prime 
\mu_{\lambda}({\bf x},{\bf x}^\prime)  J^{(\lambda)}_{N}({\bf n},{\bf x}^\prime) \ . 
\end{equation}
This formula relates Jack polynomials with $N+1$ arguments to Jack polynomials with $N$ arguments 
to the same partition ${\bf n}$ $=$ $\{n_1,\ldots,n_N\}$, $n_i \in {\mathbb N}$. 

Eq.~(\ref{anderson}), Eq.~(\ref{recurs}) and Eq.~(\ref{jack}) share the same structure: An $N$--fold integral over a multivariate function $f({\bf x}^\prime)$ with $N$ (primed) arguments and an integration kernel reproduces the same function $f({\bf x})$  with $N+1$ (unprimed) arguments. 
I will embed these examples in a general framework and show that the proper generalization of the integration kernel 
has a most natural interpretation as the coordinate representation of a particle creation operator. Moreover I will construct the corresponding annihilation operator. The resulting integral representation of the annihilation operator yields interesting new integral identities, which might be useful in other contexts as well.
 
Some of the integral formulae related with the creation operator are well--known, but to my best knowledge the integration formulae of the annihilation operator are new. 

This paper focuses on spinless Fermions. However the method is not restricted to this case. Bosons and spin $1/2$ Fermions will be addressed elsewhere. 

The paper is organized as follows. The first section supplies a precise description of the problem, whose solution will be presented in Theorem \ref{theo0}. A discussion of Theorem \ref{theo0} follows. In Sec.~\ref{applications} the results of Theorem \ref{theo0} are illustrated in several applications. Proofs will be given in 
Sec.~\ref{proof}. 

\section{Statement of the result}
\label{sec2}
We consider a Hamiltonian, describing one dimensional non--relativistic spinless interacting particles with mass $1/2$ 
\begin{eqnarray}\label{res1}
\hat{H}& =& \int_\Omega \hat{\psi}^\dagger(x)\left(-\frac{d^2}{dx^2}\right)\hat{\psi}(x)dx +\nonumber\\
  &&\int_{\Omega^2} dx dx^\prime \hat{\psi}^\dagger(x)\hat{\psi}^\dagger
        (x^\prime)V(x-x^\prime)\hat{\psi}(x^\prime)\hat{\psi}(x) \ .
\end{eqnarray}
The integration domain $\Omega$ is the real axis or a compact interval. For the first choice the spectrum of $H$ will be continuous. This has the implication that the eigenvalue problem for $H$ might have no solutions in the Hilbert space 
${\mathbf L}^2$ of square integrable $C^1$--functions in ${\mathbb R}$. In this case we look for solutions in an enhanced (rigged) Hilbert space, which allows for eigenstates of $H$ which are not normalisable. In other words, we consider the Gelfand triple ${\cal D}\subset {\mathbf L}^2 \subset {\cal D}^\prime$, where ${\cal D}$ is the space of test functions in ${\mathbf L}^2$ and ${\cal D}^\prime$ is the space of distributions, dual to ${\cal D}$. In particular ${\cal D}^\prime$ includes  eigenfunctions, which behave in the asymptotic limit as plane waves (scattering solutions of the Schr{\"o}dinger equation). For this behavior in the asymptotic limit of the wave function we introduce the term scattering boundary condition (SBC).  We consider   
\begin{equation}
\Omega \ =\ \left\{\begin{matrix} \mathbb R \ , & \qquad \mbox{\rm for scattering boundary conditions (SBC) ,}\cr 
                   [0,L] \ ,&\qquad \mbox{\rm for periodic boundary conditions (PBC).} \end{matrix}\right.
\end{equation}
If we denote the one--particle vector space by $ {\cal L}\left(\Omega\right)$, we have 
\begin{equation}
{\cal L}\left(\Omega\right)\ =\ \left\{\begin{matrix}  \ {\cal D}^\prime \ ,& 
                \qquad \mbox{\rm for scattering boundary conditions (SBC) ,}\cr 
                   {\mathbf L}^2 \ ,&\qquad \mbox{\rm for periodic boundary conditions (PBC).} \end{matrix}\right.
\end{equation}
In Eq.~(\ref{res1}) $\hat{\psi}(x)$ and $\hat{\psi}^\dagger(x)$ are Fermionic creation (annihilation) operators obeying the anticommutation relation
\begin{equation}
\label{anticom1}
\{ \hat{\psi}(x), \hat{\psi}^\dagger(x^\prime)\} \ =\  \hat{\psi}(x)\hat{\psi}^\dagger(x^\prime)+ \hat{\psi}^\dagger(x^\prime)\hat{\psi}(x) \ =\ \delta(x-x^\prime) \ .
\end{equation}
For a fixed number of particles $N$, the Hamiltonian can be written in first quantization in coordinate representation as
\begin{equation}
H_N({\bf x})\ =\ -\sum_{n=1}^N\frac{\partial^2}{\partial x^2_n}+\sum_{i\neq j}V(x_i-x_j) \ ,
\label{hamiltonian}
\end{equation}
where ${\bf x}$ denotes the set of particle positions ${\bf x}$ $=$ $\{x_1,\ldots, x_N\}$. The Hamiltonian is the direct sum $H=\bigoplus_{N=0}^\infty H_N$ acting on the Fock space ${\cal H}$ $=$  $\bigoplus_{N=0}^{\infty} {\cal H}_N$,
where 
\begin{equation}
 {\cal H}_N \ =\ \{\psi({\bf x}) \in {\cal L}\left(\Omega^N\right)| \psi({\bf x}) \ 
             {\rm completely}\ {\rm antisymmetric}\} \ ,
\end{equation}
and ${\cal L}\left(\Omega^N\right)$ $=$ ${\cal L}^N\left(\Omega\right)$. 
For $\psi({\mathbf x})$ and $\phi({\mathbf x})$ $\in$ ${\cal H}_N$  we define a pairing  
by the $N$--fold integral  
\begin{equation}
\label{scalar0}
\langle \psi_N |\phi_N\rangle \ =\ 
\int_{\Omega^N}d^N[{\bf x}] \psi^*_{N}({\bf x})\phi_{N}({\bf x})\ ,
\end{equation} 
where the infinitesimal volume element $d^N[{\bf x}]$ $=$ $\prod_{n=1}^N dx_n$ 
was introduced.
 
Our goal is to map the Hamiltonian as given in Eq.~(\ref{res1}) onto the quadratic form
\begin{equation}\label{hamiltonianq}
\hat{H} =\ \left\{\begin{matrix} \D \sum_k \epsilon(k) \hat{a}_k^\dagger \hat{a}_k & \quad ({\rm PBC})\cr 
             \D\int_k dk \epsilon(k) \hat{a}_k^\dagger \hat{a}_k & \quad ({\rm SBC}) \ ,          
            \end{matrix}\right.
\end{equation}
where $\hat{a}_k$ ($\hat{a}_k^\dagger$) annihilates (creates) a particle with quasimomentum $k$. In this form the interacting nature of the particles is hidden in the sum over the quasimomenta $k$. This sum runs over all allowed $k$ values. Whereas for SBC all real $k$ values are allowed, for PBC only special values of $k$ are allowed, corresponding to the quantisation of the theory. As we will see, interaction becomes manifest in a quantisation condition, which differs from the one for free particles.
      
The Hamiltonian (\ref{hamiltonianq}) acts on the complete Fock space and is 
block diagonal in the basis of eigenstates of the particle number operator
\begin{equation}\label{numberop}
\hat{N} =\ \left\{\begin{matrix} \D \sum_k \hat{a}_k^\dagger \hat{a}_k & \quad ({\rm PBC})\cr 
             \D\int_k dk \hat{a}_k^\dagger \hat{a}_k & \quad ({\rm SBC}) \  .         
            \end{matrix}\right. 
\end{equation}
In contrast, the creation and annihilation operators define a map which does not conserve particle number
\begin{eqnarray}\label{res4}
\hat{a}_k: {\cal H}_N \to {\cal H}_{N-1}\ &,&\  \hat{a}_k|\psi_N\rangle \in  {\cal H}_{N-1} \ , \nonumber\\ 
\hat{a}_k^\dagger: {\cal H}_N \to {\cal H}_{N+1}\ &,&\  \hat{a}_k^\dagger|\psi_N\rangle \in  {\cal H}_{N+1} \ .
\end{eqnarray}
This mapping reads in configuration space
\begin{eqnarray}\label{res5}
\psi_{N+1}({\bf x})\ &=&\ \int_{\Omega^N} d^N[{\bf x}^\prime] \bar{a}_k^\dagger({\bf x},{\bf x}^\prime)\psi_{N}({\bf x}^\prime)\ ,\nonumber\\
\psi_{N-1}({\bf x})\ &=&\ \int_{\Omega^N} d^N[{\bf x}^\prime] \bar{a}_k({\bf x},{\bf x}^\prime)\psi_{N}({\bf x}^\prime) \ .
\end{eqnarray} 
Thus, in configuration space the creation operator is an integral operator whose kernel is a complex valued function of two 
sets of coordinates ${\bf x}$ $=$ $\{x_1,\ldots, x_{N+1}\}$ and ${\bf x}^\prime$ $=$ $\{x^\prime_1,\ldots, x^\prime_N\}$. We call $\bar{a}_k^\dagger({\bf x},{\bf x}^\prime)$  {\em (antisymmetric) creation function}. The 
annihilation operator is an integral operator whose kernel $\bar{a}_k({\bf x},{\bf x}^\prime)$ is a complex valued function of ${\bf x}$ $=$ $\{x_1,\ldots, x_{N-1}\}$ and ${\bf x}^\prime$ $=$ $\{x^\prime_1,\ldots, x^\prime_N\}$. We call  $\bar{a}_k({\bf x},{\bf x}^\prime)$ {\em (antisymmetric) annihilation function}.

The creation and annihilation operators are defined by the basis independent commutator relation
\begin{eqnarray}\label{res6}
[\hat{H},\hat{a}_k]& =& -\epsilon(k) \hat{a}_k \nonumber\\
{[\hat{H},\hat{a}_k^\dagger]}& =& \epsilon(k) \hat{a}_k^\dagger  \ .
\end{eqnarray}
They read in coordinate representation 
\begin{eqnarray}
\label{res6coord1}
0&=& \left[H_{N-1}({\bf x})+\epsilon(k)\right] 
\int_{\Omega^N} d^N[{\bf x}^\prime] \bar{a}_k({\bf x},{\bf x}^\prime)\psi_{N}({\bf x}^\prime)\nonumber\\
 &&\qquad\qquad - \int_{\Omega^N} d^N[{\bf x}^\prime] 
 \bar{a}_k({\bf x},{\bf x}^\prime) H_N({\bf x}^\prime) \psi_{N}({\bf x}^\prime) \ ,\nonumber\\
0 &=& \left[H_{N+1}({\bf x})-\epsilon(k)\right] 
\int_{\Omega^N} d^N[{\bf x}^\prime] \bar{a}_k^\dagger({\bf x},{\bf x}^\prime)\psi_{N}({\bf x}^\prime)-\nonumber \\
&&\qquad\qquad \int_{\Omega^N} d^N[{\bf x}^\prime] \bar{a}_k^\dagger({\bf x},{\bf x}^\prime) H_N({\bf x}^\prime) 
   \psi_{N}({\bf x}^\prime)  \ .
\end{eqnarray}
This translates into a set of partial differential equations
for the creation functions $\bar{a}_k^\dagger({\bf x},{\bf x}^\prime)$ and the annihilation functions 
$\bar{a}_k({\bf x},{\bf x}^\prime)$ 
\begin{eqnarray}\label{res6coord}
\left[H_{N-1}({\bf x})- H_{N}({\bf x}^\prime)\right] \bar{a}_k({\bf x},{\bf x}^\prime)& 
                                  =& -\epsilon(k) \bar{a}_k({\bf x},{\bf x}^\prime) \ ,\nonumber\\ 
\left[H_{N+1}({\bf x})- H_{N}({\bf x}^\prime)\right] \bar{a}_k^\dagger({\bf x},{\bf x}^\prime)& 
                                  =& \epsilon(k) \bar{a}_k^\dagger({\bf x},{\bf x}^\prime) \ .
\end{eqnarray}
With a set of operators fulfilling the commutator relations \eqref{res6}  simultaneous
eigenfunctions of the Hamiltonians Eq.~(\ref{hamiltonianq}) and Eq.~(\ref{hamiltonian}) to 
the eigenvalue $E=\sum_{n=1}^N \epsilon(k_n)$ can be constructed 
by 
\begin{equation}\label{eigenfunctions}
|\psi_N({\bf k})\rangle \ =\ \prod_{n=1}^N \hat{a}_{k_n}^\dagger |0\rangle\ .
\end{equation}
Since two operators which have the same eigenfunctions to the same eigenvalue are equal, the conditions specified in Eqs.~(\ref{res6}) to \eqref{res6coord} are sufficient to prove that the two Hamiltonians in the forms  
given in Eq.~(\ref{res1}) and in Eq.~(\ref{hamiltonianq}) are equal up to a basis rotation. 

Since we are dealing with Fermions, we require the wave function $\langle {\bf x}|\psi_N({\bf k})\rangle \in {\cal H}_N$ to be completely antisymmetric in two sets of arguments ${\bf k}=\{k_1,\ldots,k_N\}$ and ${\bf x}=\{x_1,\ldots,x_N\}$, thus the quasiparticle creation and annihilation operators have to obey the Fermionic anticommutation rules
\begin{eqnarray}
\label{anticomm}
\{\hat{a}_k,\hat{a}_{k^\prime}^\dagger\} &=& \left\{\begin{matrix}\D
              \delta(k-k^\prime)\ ,&\qquad \mbox{\rm SBC ,}\cr 
             \D \delta_{k, k^\prime}\ , &\qquad \mbox{\rm PBC .}
	   \end{matrix}\right. 
\end{eqnarray}

From the configuration space representation (\ref{res5}) it is seen that creation function and annihilation function have to be completely antisymmetric in both sets of arguments. Since $\psi_N({\bf x}^\prime)$ is completely antisymmetric, any symmetric part of $\bar{a}_k^\dagger({\bf x},{\bf x}^\prime)$ in the primed set of arguments does not contribute to the integral. On the other hand antisymmetry of $\psi_{N+1}({\bf x})$ requires  antisymmetry in the unprimed set of arguments, as well.

We introduce for SBC the {\em statistical functions} $I^\dagger_N(\mathbf{x},\mathbf{x}^\prime)$ and  $I_N(\mathbf{x},\mathbf{x}^\prime)$ as follows
\begin{eqnarray}
\label{stat1}
I^\dagger_N(\mathbf{x},\mathbf{x}^\prime) &=&  \frac{2^{-(N+1)}}{(N+1)!}
                                   \det\left[\begin{matrix}\D 
                          \sgn(x_n-x^\prime_m),1\end{matrix}\right]_{\genfrac{}{}{0pt}{}{n=1,\ldots N+1}{m=1,\ldots N}}\nonumber  \\
I_N(\mathbf{x},\mathbf{x}^\prime) &= &  \frac{2^{-N}}{N!}\det
  \left[\begin{matrix}\D \sgn(x^\prime_m-x_n),1\end{matrix}\right]_{\genfrac{}{}{0pt}{}{n=1,\ldots N-1}{m=1,\ldots N}}          
\end{eqnarray}
In order to define statistical functions for PBC we introduce the sawtooth function
\begin{equation}
[x] \ =\ x- n L\ , \qquad n = {\rm max}(m\in {\mathbb Z}| m\leq x) \ .
\end{equation}
Then for PBC
\begin{eqnarray}
\label{stat2}
I^\dagger_N(\mathbf{x},\mathbf{x}^\prime) &=&  \frac{2^{-(N+1)}}{(N+1)!}
                                   \det\left[\begin{matrix}\D \sgn([x_n]-[x^\prime_m]),1\end{matrix}\right]_{\genfrac{}{}{0pt}{}{n=1,\ldots N+1}{m=1,\ldots N}}\nonumber  \\
I_N(\mathbf{x},\mathbf{x}^\prime) &= &  \frac{2^{-N}}{N!}\det
  \left[\begin{matrix}\D \sgn([x^\prime_m]-[x_n]),1\end{matrix}\right]_{\genfrac{}{}{0pt}{}{n=1,\ldots N-1}{m=1,\ldots N}} \ .
\end{eqnarray}
We now write the antisymmetric creation and annihilation functions as
 \begin{eqnarray}
\label{stat3}
\bar{a}^\dagger_k(\mathbf{x},\mathbf{x}^\prime) &=& I^\dagger_N(\mathbf{x},\mathbf{x}^\prime)a^\dagger_k(\mathbf{x},\mathbf{x}^\prime)\ ,\nonumber\\
\bar{a}_k(\mathbf{x},\mathbf{x}^\prime) &=& I_N(\mathbf{x},\mathbf{x}^\prime)a_k(\mathbf{x},\mathbf{x}^\prime)\ ,
\end{eqnarray}
where the creation (annihilation) functions without bar are now symmetric in both sets of arguments. The symmetric creation (annihilation) functions are required to be solutions to the same partial differential equation as the antisymmetric ones \eqref{res6coord}. In the form \eqref{stat3} the creation (annihilation) function has been separated into a dynamical factor $a_k^\dagger$ or $a_k$ and a factor $I_N^\dagger$ ($I_N$) which keeps track of the particle statistics. For this reason we call it statistical function.      

We introduce the following notation: $\psi_n({\bf k},{\bf x})$ denotes always a function with $n\in {\mathbb N}$ arguments $\{k_1,\ldots,k_n\}$ and with $n$ arguments $\{x_1,\ldots,x_n\}$.
 
We now state the main result as a theorem. It states a sufficient condition on the interaction potential $V(x)$ 
for the existence of the above defined creation and annihilation operators and gives an explicit construction of the eigenstates.
\begin{theorem}[Spinless Fermions]\label{theo0}
\begin{enumerate}
\item \label{item1} Let $f(x)$ be an antisymmetric function satisfying the condition
\begin{eqnarray}\label{theo1}
f(x)f(y)+ f(x)f(z)+f(y)f(z)& = & {\rm const.} \ , \nonumber \\
x+y+z &=& 0
\end{eqnarray}
and $F(x)=\int^x dx^\prime f(x^\prime)$, such that $F(x)$ $=$ $F(-x)$, then symmetric annihilation functions
 $a_k({\bf x},{\bf x}^\prime)$ and creation functions $a_k^\dagger({\bf x},{\bf x}^\prime)$ 
satisfying the partial differential equation (\ref{res6coord}) are given by
\begin{eqnarray}\label{theo3}
a_k({\bf x},{\bf x}^\prime)& = & \exp\left[-\sum_{n< m}^{N-1} F(x_n-x_m)+\sum_{n, m} F(x_n-x^\prime_m)-\right.\nonumber\\
&&\ \left.\sum_{n< m}^N F(x^\prime_n-x^\prime_m)+
ik\left(\sum_{n=1}^{N-1} x_n-\sum_{m=1}^{N}x_m^\prime\right)\right]\ ,\nonumber\\
a_k^\dagger({\bf x},{\bf x}^\prime) &=& \ 
\exp\left[-\sum_{n< m}^{N+1} F(x_n-x_m)+\sum_{n, m} F(x_n-x^\prime_m)-\right.\nonumber\\
&&\ \left. \sum_{n< m}^NF(x^\prime_n-x^\prime_m)+
ik\left(\sum_{n=1}^{N+1} x_n-\sum_{m=1}^{N}x_m^\prime\right)\right] \ .
\end{eqnarray}
\item \label{item2}
The interaction potential in the Hamiltonian (\ref{hamiltonian}) is related to the
function $f$  by
\begin{equation}\label{potential}
V(x)\ =\ f^2(x)-f^\prime(x)+ {\rm const.}
\end{equation}
\item \label{item3} With the complex functions $a_k({\bf x},{\bf x}^\prime)$ and $a_k^\dagger({\bf x},{\bf x}^\prime)$ defined in Eq.~(\ref{theo3}) eigenfunctions $\psi_{N\pm 1}({\bf k},{\bf x})$ to the $N\pm 1$ particle Hamiltonian Eq.~(\ref{hamiltonian})
as
\begin{eqnarray}
\label{recursiona1}
\psi_{N+1}({\bf k},{\bf x}) &=& \langle {\bf x}|\hat{a}_{k_{N+1}}^\dagger|\psi_N({\bf k})\rangle \nonumber\\
                             &=&  \frac{C_{N}({\bf k})}{\sqrt{N+1}} \int_{\Omega^N} d^N[{\bf x}^\prime]
			           I_N^\dagger({\bf x},{\bf x}^\prime)
			           a_{k_{N+1}}^\dagger({\bf x},{\bf x}^\prime)\psi_{N}({\bf k},{\bf x}^\prime)\ ,\\ 
\label{recursionb1}
\psi_{N-1}({\bf k},{\bf x}) &=& \langle {\bf x}|\hat{a}_{k_N}|\psi_N({\bf k)}\rangle \nonumber\\
                                      &= &  R \sqrt{N}C_{N-1}({\bf k})
				       \int_{\Omega^N}d^N[{\bf x}^\prime] I_N({\bf x},{\bf x}^\prime) 
				       a_{k_{N}}({\bf x},{\bf x}^\prime)\psi_{N}({\bf k},{\bf x}^\prime)\ ,
\end{eqnarray} 
where $R= (2\pi)^{-1}$ for SBC and $R=L^{-1}$ for PBC. 
The normalization constant $C_N({\bf k})$ is coordinate independent. For the most 
important potentials the explicit value will be given below (see Proposition \ref{prop1}).
\item \label{item4}
The dispersion relation of the quasimomenta is quadratic:
\begin{equation}
\label{dispersion}
\epsilon(k)= k^2 \ .
\end{equation}
The functions (\ref{recursiona1}) and (\ref{recursionb1}) are eigenfunctions to the $N\pm1$ particle Hamiltonians \eqref{hamiltonian} and to the center of mass momentum operator 
\begin{equation}
\label{com}
P_{N\pm1}({\bf x}) \ =\ -i\sum_{n=1}^{N\pm 1} \frac{\partial}{\partial x_n} \ , 
\end{equation}
such that
\begin{eqnarray}
 H_{N\pm1}({\bf x}) \psi_{N\pm1}({\bf k},{\bf x}) & = &
          \left( \sum_{n=1}^{N\pm 1} k_n^2\right) \psi_{N\pm1}({\bf k},{\bf x})\ , \\
 P_{N\pm 1}({\bf x})  \psi_{N\pm1}({\bf k},{\bf x})&=& \left(\sum_{n=1}^{N\pm 1} k_n\right) \psi_{N\pm1}({\bf k},{\bf x})\ .
\end{eqnarray}
\item \label{item5} 
With respect to the pairing (\ref{scalar0}) the orthogonality relation
\begin{equation}
\label{scalar1}
\langle \psi_N({\bf k}^\prime) |\psi_N({\bf k})\rangle \ \propto\ \left\{
 \D\begin{array}{cc}\D\det [2\pi \delta(k^\prime_i-k_j)]_{1\leq i,j\leq N} \ ,& {\rm SBC}\cr
 \D\det \left[ L \delta_{k^\prime_i, k_j}\right]_{1\leq i,j\leq N} \ ,& {\rm PBC}.\end{array}\right.
\end{equation}
holds.  
\end{enumerate}
\end{theorem}
The proof of Theorem \ref{theo0} will be given in Sec.~\ref{proof}. Here we discuss some of its consequences. 

\paragraph{Condition on the potential}
The functional equation (\ref{theo1}) is a special case of the functional equation 
\begin{eqnarray}
f(x)f(y)+f(x)f(z)+f(y)f(z) & = & W(x)+W(y)+W(z)\nonumber\\
x+y+z & = &0  \ , \label{Frobstickel}
\end{eqnarray}
which was found by Sutherland \cite{sut71,sut75} to be the necessary condition for a product wave function 
\begin{equation}\label{product}
 \psi_N({\bf 0},{\bf x}) \ =\ \prod^N_{i<j}\exp[F(x_i-x_j)]
\end{equation}
 to be an eigenfunction to eigenvalue zero of an $N$--body Hamiltonian with two--body interaction only. 
The general solution of \eqref{Frobstickel} was found by Calogero \cite{cal75a,cal75b}. It is given by
\begin{equation}
\label{solution}
f(x)\ =\ \lambda \zeta(x|\omega,\omega^\prime) + \beta x \ ,
\end{equation}
where $\zeta(x|\omega,\omega^\prime)$ is the Weierstrass $\zeta$--function \cite{abr72} 
and $\omega,\omega^\prime$, ordered by ${\rm Re}(\omega)\geq {\rm Re}(\omega^\prime)$ are the two periods of the corresponding Weierstrass elliptic function ${\cal P}$. 
The right hand side of \eqref{Frobstickel} is determined by the Frobenius--Stickelberger equation 
for the Weierstrass $\zeta$--function \cite{fro80} 
\begin{eqnarray}
\label{Frobstickel1}
\left(\zeta(x)+ \zeta(y) +\zeta(z)\right)^2 & = & - \zeta^\prime(x)-\zeta^\prime(y)-\zeta^\prime(z) \nonumber\\
                       x+y+z  & = & 0  \ ,
\end{eqnarray}
to 
\begin{eqnarray}
W(x) & = & - \frac{1}{2}\left(\lambda f^\prime(x) + f^2(x) -\lambda\beta\right). 
\end{eqnarray}
The condition $3 W(x)$ $=$ ${\rm const.}$ therefore also requires 
\begin{equation}
\label{solution1}
-\frac{3}{2}\left(\lambda f^\prime(x)+f^2(x)-\lambda \beta\right) \ =\  {\rm const.} \ .
\end{equation}
Eq.~(\ref{solution1}) can be integrated. We express the constant as
\begin{equation}
\label{solution1a}
{\rm const.} = \frac{3\lambda}{2}\left(\beta-z^2\lambda\right)
\end{equation}
with $z\in {\mathbb C}$. The most general simultaneous 
solution to Eq.~(\ref{solution1}) and Eq.~(\ref{theo1}) can then be written as
\begin{equation}
\label{gensol}
f(x) \ =\   z\lambda \coth\left(zx-\kappa \right) \ ,
\end{equation}
with  arbitrary complex parameters $\kappa$, $z$ and $\lambda$. Requirement of a real potential restricts the values of $z=a+ib$,  $a, b \in {\mathbb R}$ to be either real $z=a$ or purely imaginary $z=ib$ and $\lambda$ to be real. 
A series of potentials can be derived by taking various limits. The most important  cases are listed
in Table \ref{table1}. They are the interactions of the trigonometric (I), rational (II) and hyperbolic (III)  Calogero--Sutherland--Moser (CMS) type.
Remarkably also the sign--function respectively the $\delta$--distribution are obtained from the shifted hyperbolic CMS Hamiltonian in the limit $a\to \infty$, $\lambda\to 0$ with $a\lambda= c $ finite. For sake of completeness in Table \ref{table2} for these potentials the values of the parameters $a$, $b$, $\kappa$, ${\rm const.}$ and of the corresponding periods $\omega,\omega^\prime$ are listed.
\begin{table}[tbp] \centering
\begin{tabular}
[c]{c|c|c|c|c}
  type          &  name          &   $f(x)$         & $V(x)$   &  $F(x)$      \\ \hline
 (I)            &  trigonometric CMS    & $\D \lambda b \cot(b x)$          
              & \parbox{2cm}{\vspace{0.5em}$\D \frac{b^2\lambda(\lambda+1)}{\sin^2(x b)}$}  
	  & $\D\lambda\ln|\sin(bx)|$  
	  \\ \hline
 (II) & rational CMS&       $\frac{\displaystyle\lambda}{\displaystyle x}$ 
    &   \parbox{2cm}{\vspace{0.5em} $\lambda(\lambda+1)\frac{1}{\displaystyle x^2}$}
 & $\lambda\ln |x|$    \\ \hline
 (III) & hyperbolic CMS         & $a \lambda\coth(ax)$          
    &   \parbox{2cm}{\vspace{0.5em}$ \frac{\displaystyle a^2\lambda(\lambda+1)}{\displaystyle\sinh^2(x a)} $} & 
  $\displaystyle\lambda\ln|\sinh(ax)|$   \\ \hline
  (IV) & Morse potential        & $a \lambda\tanh(ax)$          
    &   \parbox{2cm}{\vspace{0.5em}$\D -\frac{ a^2\lambda(\lambda+1)}{\cosh^2(x a)}$} & 
  $\displaystyle\lambda\ln|\cosh(ax)|$   \\ \hline 
  (V) &$\delta$--interaction   & $\D c \ \sgn(x)$ & \parbox{1cm}{\vspace{0.5em}$\D -2 c\delta(x)$}  & $\D c |x|$  
\end{tabular}
\caption{\label{table1} List of real interaction potentials for which creation and annihilation operators can be constructed from the general solution Eq.{\ref{gensol}}}. 
\end{table}%

\begin{table}\centering
\begin{tabular}
[c]{c|c|c|c|c|c|c|c}
  type          &  a          &   b         &  $\kappa$   & ${\rm const.}$ & $\beta$   &  $\omega$ & $\omega^\prime$ \\  \hline

\parbox{1cm}{\vspace{0.7em}(I)} &  $0$ &   free & $0$ & $b^2\lambda^2$ &  $-\lambda b^2/3$  &  $\pi/2b$ &  $i\infty$       \\  \hline
\parbox{1cm}{\vspace{0.7em}(II)}&  $0$ &   $0$  & $0$ &      $0$       &      $0$     &  $\infty$ &   $i\infty$      \\ \hline
\parbox{1cm}{\vspace{0.7em}(III)}&  free&  $0$ &   $0$  &   $-a^2\lambda^2$&  $\lambda a^2/3$  &$\infty$ & $\pi/2ia$       \\  \hline
\parbox{1cm}{\vspace{0.7em}(IV)}  & free& $0$  &  $i\pi/2$   & $-a^2\lambda^2$& $\lambda a^2/3$ &$\infty$ & $\pi/2ia$  \\  \hline 
  \parbox{1cm}{\vspace{0.7em}(V)} &\parbox{1cm}{\vspace{0.7em} $\genfrac{}{}{0pt}{0}{a\to\infty}{a\lambda\ {\rm finite}}$}&   $0$
       &    $0$      &    $-c^2 = -\lambda^2a^2$ 
 & \parbox{1cm}{\vspace{0.7em} $\genfrac{}{}{0pt}{0}{\beta\to\infty}{\beta \lambda\ {\rm finite}}$}         &  $\infty$ &   $i\infty$ 
\end{tabular}
\caption{\label{table2} Values of the parameters $a$, $b$ and $\kappa$ for the potentials described in Table \ref{table1}. 
The corresponding values for $\omega,\omega^\prime$ and $\beta$ defined through Eq.~(\ref{solution}) are stated as well}.  
\end{table}

\paragraph{Structure of creation and annihilation functions}

The creation (annihilation) functions as defined in Eq.~(\ref{theo3}) are strictly speaking also functions of the particle number $N$. We suppressed this obvious $N$ dependence in order to unburden notation. However, to describe the relation 
between $a_k$ and $a_k^\dagger$ it is useful to indicate the $N$ dependence by the symbols $a_k^{(N)}({\bf x},{\bf x}^\prime)$ and  by $a_k^{\dagger(N)}({\bf x},{\bf x}^\prime)$. Then we have 
\begin{equation}
\label{creaanihl}
{a_k^{(N+1)}}^*({\bf x}^\prime,{\bf x}) \ =\ {a_k^{\dagger(N)}}({\bf x},{\bf x}^\prime) \ .
\end{equation}
The general structure of the creation (annihilation) function $a_{k}^{\dagger}({\bf x},{\bf x}^\prime)$ ($a_{k}({\bf x},{\bf x}^\prime)$) factorizes into a $k$ independent part $a_{0}^{\dagger}({\bf x},{\bf x}^\prime)$ ($a_{0}({\bf x},{\bf x}^\prime)$) and a $k$ dependent part.
The $k$ dependent part is for all types of potentials a product of plane waves of the primed and the unprimed coordinates. They may be considered arbitrary eigenfunctions of the non--interaction $V(x)=0$ many--body Hamiltonian. On the other hand, as mentioned above, the basic ingredient of the $k$--independent part is the product function 
$\psi_N({\bf 0},{\bf x})=\prod_{i<j}\exp[F(x_i-x_j)]$, which is an eigenfunction of the interacting system to eigenvalue zero. 

Using this general structure one might try to extend Theorem \ref{theo0} to systems, where translation invariance is broken by an external potential, but whose ground state wave function can yet be written in the product form of Eq.~(\ref{product}). 
It was shown \cite{ino84,kop00} that allowing for an external potential enhances considerably the number of systems with a factorizing ground state. For most of these systems only the ground state is known exactly. Whether or not some or all exited states can be constructed exactly is an open question.   

\paragraph{Statistical functions and Integration boundaries}

The statistical functions (\ref{stat1}) and (\ref{stat2}) are conceptually important. They give a clue how to modify the method to particles with Bosonic or (more generally) anionic statistics. For practical purposes they are not very convenient. We define integration domains $I^{(N)}_{\rm in}$ and $I^{(N)}_{\rm out}$ as
\begin{eqnarray}
\label{intdomain}
\int_{I^{(N)}_{\rm in}} d^N[{\bf x}^\prime]\left(\ldots\right) &=& \int_{x_2}^{x_1}dx_1^\prime\int_{x_3}^{x_2}dx_2^\prime\ldots\int_{x_{N+1}}^{x_{N}}dx_{N}^\prime\left(\ldots\right) \nonumber\\
\int_{I^{(N)}_{\rm out}} d^N[{\bf x}^\prime]\left(\ldots\right) &=& \int^{X_0}_{x_1}dx_1^\prime\int_{x_2}^{x_1}dx_2^\prime\ldots\int_{X_1}^{x_{N-1}}dx_{N}^\prime\left(\ldots\right)\ ,
\end{eqnarray}
where $X_0\equiv +\infty$ and $X_1\equiv -\infty$ for SBC, and $X_0\equiv L$ $X_1\equiv 0$ for PBC. Then
\begin{eqnarray}
\int_{\Omega^N} d^N[{\bf x}^\prime] I^\dagger_N({\bf x},{\bf x}^\prime)\chi_N({\bf x}^\prime)&=& \int_{I^{(N)}_{\rm in}} d^N[{\bf x}^\prime]\chi_N({\bf x}^\prime) \label{bound1}\\
\int_{\Omega^N} d^N[{\bf x}^\prime] I_N({\bf x},{\bf x}^\prime)\chi_N({\bf x}^\prime)&=& \int_{I^{(N)}_{\rm out}} 
                                  d^N[{\bf x}^\prime]\chi_N({\bf x}^\prime) \label{bound2}
\end{eqnarray}
for an arbitrary antisymmetric test function $\chi_N({\bf x}^\prime)\in {\cal D}\subset {\mathcal H}_N$. Eqs.~\eqref{bound1} 
and \eqref{bound2} are proven by a direct calculation using $\theta(x)\ =\ \frac{1}{2}(\sgn(x)+1)$ and properties of the determinant. Using the integration domains, introduced above, the recursion relations \eqref{recursiona1} and \eqref{recursionb1} read   
\begin{eqnarray}
\label{recursiona}
\psi_{N+1}({\bf k},{\bf x})  &=&  \frac{C_{N}({\bf k})}{\sqrt{N+1}} \int_{I^{(N)}_{\rm in}} d^N[{\bf x}^\prime]
			                            a_{k_{N+1}}^\dagger({\bf x},{\bf x}^\prime)\psi_{N}({\bf k},{\bf x}^\prime)\ ,\\ 
\label{recursionb}
\psi_{N-1}({\bf k},{\bf x}) &= &  R \sqrt{N}C_{N-1}({\bf k})
				       \int_{I^{(N)}_{\rm out}}d^N[{\bf x}^\prime] a_{k_{N}}({\bf x},{\bf x}^\prime)\psi_{N}({\bf k},
				              {\bf x}^\prime)\ .
\end{eqnarray} 
This form turns out to be more convenient for calculations.
 
\paragraph{Periodicity}
For periodic boundary conditions the newly created wave functions $\psi_{N+1}({\bf k},{\bf x})$ and 
$\psi_{N-1}({\bf k},{\bf x})$ must be periodic with period $L$ in all arguments. For PBC the statistical functions $I_N^\dagger$ and $I_N$ are periodic as well. This yields the following restrictive condition onto the form of the symmetric creation (annihilation) functions. 

\begin{proposition}[Periodicity]\label{periodicity}
For PBC the symmetric creation (annihilation) function $a_k^\dagger({\bf x},{\bf x}^\prime)$ ($a_k({\bf x},{\bf x}^\prime)$) 
itself must be periodic in $x$ with period $L$. 
\end{proposition}
Proposition \ref{periodicity} essentially fixes the 
boundary condition for all potentials in Table \ref{table1}.
Potential (I) is periodic with period $\pi/b$. Therefore the asymptotic regime is never reached and only periodic boundary conditions $L= \pi/b$ are allowed for this potential.
By Prop.~\ref{periodicity} for potentials of type (II) to (IV) only scattering boundary conditions (SBC) are allowed\footnote{ Continuing potential (II) periodically leads to potential (I).}. 
Only for particles with $\delta$--interaction both boundary conditions can be imposed.

\paragraph{N-particle wave function}
As a corollary to Theorem \ref{theo0} every $N$--particle wave function can be written as a 
$N(N-1)/2$--fold integral as follows
\begin{eqnarray}
\label{Nfoldint}
\psi_{N}({\bf k},{\bf x}) &=& \prod_{n=1}^{N-1}\left( \frac{C_n({\bf k})}{\sqrt{n+1}}\int_{I^{(n)}_{\rm in}}d^n[{\bf x}^{(n)}] 
               a_{k_n}^\dagger({\bf x}^{(n+1)},{\bf x}^{(n)})\right)  \ .
\end{eqnarray}
The coordinate sets ${\bf x}$ and ${\bf x}^{(N)}$ are identified. Eq.~(\ref{Nfoldint}) is obtained by iterating Eq.~(\ref{recursiona}).

\paragraph{Analyticity at the Boundaries}
For the integrals (\ref{recursiona}) and (\ref{recursionb}) to exist the creation and annihilation functions can have at most an integrable singularity in their integration domain. For potentials (I) to (III) this yields a restriction for the coupling parameter $\lambda$ to the range $\lambda \in (-1,\infty)$. This restriction is consistent with the well known fact that for CMS--Hamiltonians the minimal value of the coupling constant $g=\lambda(\lambda+1)$ is $g=-1/4$ \cite{ols83}. For these potentials the behavior of $\psi_N({\bf k},{\bf x})$ when two particles come close to each other can also be extracted directly from the creation and annihilation functions. The wave function vanishes with a typical power
\begin{equation}
 \lim_{x_i\to x_{i+1}}\psi_N({\bf k},{\bf x}) \ \propto \  (x_i-x_{i+1})^{\lambda+1}+ {\cal O}((x_i-x_{i+1})^{\lambda+2})\ .
\end{equation} 
This shows again that $\lambda \geq -1$ must be imposed.
    
\paragraph{Normalisation}
For potentials (II)--(V) with SBC the wave function is not normalisable. The constant $C_N({\bf k})$ can be evaluated by requiring that in the asymptotic regime, i.~e.~in the regime where all distances $x_i-x_j$, $1\leq i,j\leq N$ are large the wave function $\psi_{N}({\bf k},{\bf x})$ obtains the form of a scattering wave solution 
\begin{equation}
\label{asymp}
\psi^{\rm (asym)} _{N}({\bf k},{\bf x}) \ =\ \frac{1}{\sqrt{N!}}
                                \sum_{\omega \in S^{N}} S_\omega({\bf k})e^{i\sum_{n=1}^{N}x_n k_{\omega(n)}}\ ,
\end{equation}  
where $S^N$ is the permutation group and $S_\omega({\bf k})$ is the momentum dependent $N$--body scattering matrix.
For potential (I) the wave--function is normalisable and the calculation is different. The result for all cases can be summarized as follows.
\begin{proposition}
\label{prop1}
Depending on the potentials listed in Table \ref{table1} the  $k$--dependent normalisation constant $C_N({\bf k})$ is given by
\begin{eqnarray}
\label{norm1}
C_N({\bf k}) & = &\left\{\begin{matrix}
                     \D \left(\frac{L}{2\pi i}\right)^{-N}
                        \prod_{i=1}^{N}
                 \frac{(2i)^\lambda\sgn(k_i-k_{N+1})}{B\left(\frac{L}{2\pi}|k_i-k_{N+1}|,\lambda+1\right)} 
			               &\quad  \mbox{\rm for (I)\ ,}\cr
                     \D  \frac{i^{N(\lambda+1)}}{\Gamma^N(\lambda+1)} \prod_{i=1}^N|k_i-k_{N+1}|^\lambda
		        \left(k_i-k_{N+1}\right) &\quad  \mbox{\rm for (II)\ ,}\cr
                     \D (2a)^N \D \prod_{i=1}^N \frac{2^\lambda\sgn(k_i-k_{N+1})}
                     {B\left(\frac{i}{2a}|k_i-k_{N+1}|,\lambda+1\right)}    &\quad \mbox{\rm for (III),}\cr
                       \D \prod_{i=1}^N \frac{2^\lambda(i k_i-i k_{N+1})}
                      {F\left(-\lambda,\frac{i}{2a}|k_i-k_{N+1}|,\frac{i}{2a}|k_i-k_{N+1}|+1;-1\right)}  
                             &\quad \mbox{\rm for (IV)\ ,}\cr
                     \D \prod_{i=1}^N(ik_{N+1}-ik_i)    &\quad \mbox{\rm for (V)\ ,}\end{matrix}\right.  
\end{eqnarray}
\end{proposition}
where $B(x,y)$ is Euler's beta--function and $F(a,b,c;z)$ is the hypergeometric function as defined in \cite{abr72}. The proof of Prop.~\ref{prop1} is given in Sec.~\ref{AppB}.

\section{Applications}
\label{applications}
The $N(N-1)/2$ integrals in the representation Eq.~(\ref{Nfoldint}) of the exact $N$ particle state can in some cases be evaluated exactly. However, many important properties of the wave function can actually be extracted from Eq.~(\ref{Nfoldint}) without solving the integral. In the following we discuss free Fermions as well as the interaction potentials (I) to (V) of Tab.~\ref{table1}.

\subsection{Free particles}
\label{free particles} 
Since it is instructive to see, how Theorem \ref{theo0} works in the simplest case, 
we illustrate it first for free Fermions. A general $N+1$ particle state is given as the Slater determinant 
\begin{eqnarray}
\label{freefermions}
\langle{\bf x}|\psi_{N+1}({\bf k})\rangle  \ = \ \frac{1}{\sqrt{(N+1)!}}\det\left[e^{ik_nx_m}\right]_{1\leq n,m\leq N+1} &\equiv &\psi^{(0)}_{N+1}({\bf k},{\bf x})\ .
\end{eqnarray}
This result can be derived from Theorem \ref{theo0} by induction as follows:
Assume the $N$ particle states have been constructed, then we find the $N+1$ particle states by
\begin{eqnarray}
\label{freefermicreat}
\langle{\bf x}|\psi_{N+1}({\bf k})\rangle & = & \langle{\bf x}|\hat{a}_{k_{N+1}}^\dagger|\psi_{N}({\bf k})\rangle\nonumber\\
          &=& \frac{C_N({\bf k})}{\sqrt{(N+1)!}}\int_{I^{(N)}_{\rm in}} d^N[{\bf x}^\prime] 
	    a_{k_{N+1}}^\dagger({\bf x},{\bf x}^\prime)\det\left[e^{ik_nx^\prime_m}\right]_{1\leq n,m\leq N}\nonumber\\
          &=&\frac{C_N({\bf k})}{\sqrt{(N+1)!}}
	    \int_{I^{(N)}_{\rm in}} d^N[{\bf x}^\prime] 
         \det\left[e^{ik_nx^\prime_m}\right]_{1\leq n,m\leq N} \nonumber\\
	  &&\qquad \exp\left(ik_{N+1}\sum_{n=1}^{N+1} x_n-ik_{N+1}\sum_{m=1}^{N}
	   x^\prime_m\right) \ .
\end{eqnarray} 
With the integration domain $I^{(N)}_{\rm in}$ given by Eq.~(\ref{intdomain}) 
the integral can be performed yielding
again a determinant
\begin{equation}
\langle{\bf x}|\psi_{N+1}({\bf k})\rangle \ =\ \frac{1}{\sqrt{(N+1)!}}\frac{C_N({\bf k})}{\prod_{n=1}^{N}(ik_n-ik_{N+1})}
                         \det\left[e^{ik_nx_m}\right]_{1\leq n,m\leq N+1 }\ ,
\end{equation}
which is the desired result (compare with Prop.~\ref{prop1}). 
The action of $\hat{a}_{k}$ on an $N$ particle state is given by
\begin{eqnarray}
\label{freefermannihil}
\langle{\bf x}|\psi_{N-1}({\bf k})\rangle & = & \langle{\bf x}|\hat{a}_{k}|\psi_{N}({\bf k})\rangle\nonumber\\
          &=&\frac{R C_{N-1}({\bf k})}{\sqrt{(N-1)!}}
         \int_{I^{(N)}_{\rm out}} d^N[{\bf x}^\prime] a_{k}({\bf x},{\bf x}^\prime)
	  \det\left[e^{ik_nx^\prime_m}\right]_{1\leq n,m \leq N }\nonumber\\
          &=&\frac{R C_{N-1}({\bf k})}{\sqrt{(N-1)!}}\int_{I^{(N)}_{\rm out}} d^N[{\bf x}^\prime] 
	  \det\left[e^{ik_nx^\prime_m}\right]_{1\leq n,m\leq N}\nonumber\\
	  &&\qquad \exp\left(ik\sum_{n=1}^{N-1} x_n- ik\sum_{m=1}^{N} x^\prime_m\right) \ .
\end{eqnarray}
If we choose SBC we have to equip the quasimomentum $k$ with a positive imaginary increment $i\epsilon$ 
for the $x_1^\prime$ integration and with a negative increment for the $x_N^\prime$ integration. Then the integrals are convergent
and yield again a determinant
 \begin{eqnarray}
\label{freefermannihil2}
&&\langle{\bf x}|\psi_{N-1}({\bf k})\rangle \ =\frac{1}{2\pi}\frac{C_{N-1}({\bf k})}{\sqrt{(N-1)!}}
           \left|\begin{matrix}\D2\pi \delta(k-k_1)
            &\D\frac{e^{ik_1x_1}} {ik_1-ik}&\ldots&\D\frac{e^{ik_1x_{N-1}}}{ik_1-ik}\cr
	                        \D    \vdots        &\vdots                  &\ddots&\vdots   \cr
				 \D 2\pi \delta(k-k_N)&\D \frac{e^{ik_Nx_1}}
                             {ik_N-ik}&\ldots&\D\frac{e^{ik_Nx_{N-1}}}{ik_N-ik} 
                                    \end{matrix}\right| \ .
\end{eqnarray}
From Eq.~(\ref{freefermannihil2}) it follows
\begin{eqnarray}
\label{fermannihilres}
\hat{a}_k|\psi_N({\bf k})\rangle &=& 0 \ ,\ \mbox{ if $k\neq k_i$, $i=1\ldots N$}\nonumber\\
\langle{\bf x}|\hat{a}_k|\psi_N({\bf k})\rangle &=& \frac{C_{N-1}({\bf k})}{\prod_{i=1}^N (ik_i-ik)}\frac{1}{\sqrt{(N-1)!}}\nonumber\\
          &&\quad \det\left[e^{ik_nx_m}\right]_{\genfrac{}{}{0pt}{}{1\leq n\leq N-1}{1\leq n \leq N, n\neq j}}\  , 
	                                                \ {\rm for}\  k=k_j\ ,
\end{eqnarray} 
which is the desired result (compare with Prop. \ref{prop1} with $j=N$). For PBC we require $\psi_N^{(0)}({\bf k},{\bf x})$ to be a periodic function with period $L$. Therefore ${\bf k}$ $=$ $\frac{2\pi}{L}{\bf n}$, where ${\bf n}$ is a set of $N$ integers $n_i\in {\mathbb Z}$.  For the action of the creation operator nothing changes as compared to SBC. For the action of the annihilation operator we find that the $\delta$--distribution is substituted by a Kronecker--delta
\begin{equation}\label{finitedelta}
\delta(k_i-k_j)\ \to \ \frac{L}{2\pi} \delta_{k_i, k_j}\ .
\end{equation}
This completes the construction for free spinless Fermions.

\subsection{Particles with $\delta$--interaction}
\label{delta}
For spinless Fermions the $\delta$--interaction is invisible and the wave function becomes identical with the 
wave function of free Fermions as given in Eq.~\eqref{freefermions}. This result is quickly derived using Theorem \ref{theo0}. 
Using the free solution $\psi^{(0)}_N(,{\bf k},{\bf x}^\prime)$ for $\psi_N({\bf k},{\bf x}^\prime)$ in Eq.~\eqref{recursiona}, we obtain 
\begin{eqnarray}
\label{app1}
\langle{\bf x}|\psi_{N+1}({\bf k})\rangle & = & \langle{\bf x}|\hat{a}_k^\dagger|\psi_{N}({\bf k})\rangle\nonumber\\
          &=& \frac{C_N({\bf k})}{\sqrt{(N+1)!}}\int_{I^{(N)}_{\rm in}} d[{\bf x}^\prime] 
         a_k^\dagger({\bf x},{\bf x}^\prime)
	   \det\left[e^{ik_nx_m^\prime}\right]_{1\leq n,m\leq N} \ .
\end{eqnarray}
For the creation function $a_{k_{N+1}}^\dagger({\bf x},{\bf x}^\prime)$ 
we use Eq.~(\ref{theo3}) and extract $F(x)= c|x|$ from Tab.~\ref{table1} 
\begin{eqnarray}
\label{app2}
a_{k_{N+1}}^\dagger({\bf x},{\bf x}^\prime)&=& \exp\left(ik_{N+1}\sum_{n=1}^{N+1} x_n-ik_{N+1}\sum_{m=1}^{N} x^\prime_m\right) \\ 
	 && 
          \exp\left(-c\sum_{n>m}|x_n-x_m|+c \sum_{n,m}|x_n-x^\prime_m|-c \sum_{n>m}|x^\prime_n-x^\prime_m|\right) \ .\nonumber
\end{eqnarray}
We use the ordering \eqref{interlacing} of the primed and the unprimed variables implied by the boundary $I_{\rm in}^{(N)}$. With this ordering it is readily seen that the exponent in the second line of Eq.~\eqref{app2} drops out completely for any ordering of the particles. Thus Eq.~(\ref{app1}) reduces to the free particle expression Eq.~(\ref{freefermicreat}).
The corresponding result is obtained for the action of the annihilation operators.  

\subsection{ Trigonometric Calogero--Moser--Sutherland system}
\label{trig}
For the interaction potential (I) the model is called trigonometric CMS--model. As was pointed out in Sec.~\ref{sec2} the Hamiltonian has to be considered with PBC. First we recall some facts about the model. A rather comprehensive treatment of the trigonometric CMS--model is given in \cite{kur09}.

The eigenfunctions of the trigonometric CMS--models can be 
written as 
\begin{eqnarray}
\label{jack1a}
 \psi_{N}({\bf n}, {\bf z}) & = & \psi_N({\bf 0},{\bf z})\prod_{i=1}^N z_i^K J^{(\lambda)}_{N,K}({\bf n},{\bf z}) \ ,\nonumber\\
  \psi_N({\bf 0},{\bf z}) &=&   \Delta_N^{\lambda+1}({\bf z})
                                    \prod_{i=1}^N z_i^{-\frac{\lambda+1}{2}(N-1)} \ ,
 \end{eqnarray}
where the arguments $z_i$ are related to the particle positions $x_i$  by the relation 
\begin{equation}
 z_i= \exp\left(2\pi i x_i/L\right) \ .
\end{equation}
Here $\psi_{N}({\bf 0},{\bf z})$ is the ground state wave function of $H_N$ and $J^{(\lambda)}_{N,K}({\bf n},{\bf z})$ is a symmetric polynomial in $N$ variables $z_i$ labeled by a partition ${\bf n}$ $=(n_1,\ldots,n_N)$ of integers $n_1\leq n_2\leq\ldots n_N$. These polynomials are called Jack polynomials and were extensively studied \cite{mac92,sta89}. The parameter $\lambda$ is related to the parameter $\alpha$ of McDonalds book \cite{mac92} by $\lambda= 1/\alpha-1$. 
The center of mass momentum $K$ is a real parameter. Jack polynomials are defined as eigenfunctions of the operator $H_{N,K}^\prime$, which is obtained by adjunction of $H_N$ with the ground state wave function $\psi_N({\bf 0},{\bf z})$ times the Galilean boost $\prod_{i=1}^N z_i^K$
\begin{eqnarray}
H_{N,K}^\prime & = & \left(\psi_N({\bf 0},{\bf z})\prod_{i=1}^Nz_i^K\right)^{-1}\left(H_N - E_0^{(N,K)}\right)
                      \left(\psi_N({\bf 0},{\bf z})\prod_{i=1}^Nz_i^K\right)\nonumber\\
           & = &\left(\frac{2\pi}{L}\right)^2\left[\sum_{i=1}^N\left(z_i\frac{\partial}{\partial z_i}\right)^2 + 
	   \left[(\lambda+1)(N-1)+2K\right]\sum_i z_i\frac{\partial}{\partial z_i}\right.  \nonumber\\
	   &&\qquad  \left. + 2(\lambda+1)\sum_{i<j}^N\frac{z_iz_j}{z_i-z_j}
	   \left(\frac{\partial}{\partial z_i}-\frac{\partial}{\partial z_j}\right)\right] \ .
\end{eqnarray}
The ground state energy $E_0^{(N,K)}$ is given by 
\begin{equation}
E_0^{(N,K)} \ =\ \frac{1}{12}(N+1)N(N-1)(\lambda+1)^2 + N K(K+(\lambda+1)(N-1)) \ .
\end{equation}
For $K=0$ and $\lambda=0$ this is identical to the ground state energy of free spinless Fermions. 
The wave function $\psi_{N}({\bf n}, {\bf z})$ in Eq.~(\ref{jack1a}) vanishes with the power $\lambda+1$, when two particles come close to each other, but it is not antisymmetric under interchange of two particles. Rather it obtains a phase $\pi (\lambda+1)$ under the action of the permutation operator $P$
\begin{equation}
\label{exchange}
P_{nm}\psi(x_1,\ldots,x_n,\ldots,x_m,\ldots,x_N) = \psi(x_1,\ldots,x_m,\ldots,x_n,\ldots,x_N)\ , \forall n,m \ . 
\end{equation} 

For even $\lambda$  $\psi_{N}({\bf n}, {\bf z})$ is Fermionic, for odd $\lambda$ it is Bosonic. 
For arbitrary real $\lambda$ it is a wave function with anyonic statistics \cite{ha95}. 
A Fermionic wave function can always be obtained from Eq.~(\ref{jack1a}) by the substitution
\begin{equation}
 \Delta_N^{\lambda+1}({\bf z})\prod_{i=1}^N z_i^{-\frac{\lambda+1}{2}(N-1)}\ \to\ |\Delta_N({\bf z})|^{\lambda}\Delta_N({\bf z})
 \prod_{i=1}^N z_i^{-(N-1)/2}\ .
\end{equation} 
Jack polynomials depend strictly speaking also on the center of mass momentum $K$. 
In the following we fix the Galilean boost to $K$ $\equiv$ $-\frac{\lambda+1}{2}(N-1)$ and suppress the $K$ dependence of the Jack polynomial by setting \begin{equation}
J^{(\lambda)}_{N,-\frac{\lambda+1}{2}(N-1)}({\bf n},{\bf z})\ \equiv\  J^{(\lambda)}_{N}({\bf n},{\bf z})\ .
\end{equation}
We now prove three statements
\begin{proposition}\label{prop2}
The asymptotic Bethe Ansatz equation \cite{sut71b} 
\begin{equation}
\label{bethte1}
k_i\ =\ \frac{2\pi}{L}\left( I_i + \frac{\lambda}{2}\sum_{j=1}^{N+1}\sgn(k_i-k_j)\right) \ , \quad I_i\in {\mathbb Z}
\end{equation}
for the quasimomenta $k_i$ are equivalent
to the periodicity condition Prop.~\ref{periodicity} of the creation (annihilation) operator $a_k^\dagger$ ($a_k$)\ .  
\end{proposition}

\paragraph{Proof}
In the region 
\begin{equation}
x_1> x_2>\ldots> x_{N+1}
\end{equation}
the creation function $a_{k_{N+1}}^\dagger({\bf x},{\bf x}^\prime)$ can be written as 
\begin{eqnarray}
a_{k_{N+1}}^\dagger({\bf x},{\bf x}^\prime) & = & \left(\frac{\prod_{i=1}^N\prod_{j=i}^N
                                        \sin(\pi (x_i-x_j^\prime)/L)\prod_{j=1}^N
                                        \prod_{i=j+1}^{N+1}\sin(\pi (x_j^\prime-x_i)/L)}
                                       {\prod_{i<j}^{N+1}\sin(\pi (x_i-x_j)/L)
                                      \prod_{i<j}^N\sin(\pi (x_i^\prime-x_j^\prime)/L)}\right)^\lambda\nonumber\\
                                    &&\exp\left(i k_{N+1}\sum_{j=1}^{N+1} x_j -i k_{N+1}\sum_{j=1}^{N} x_j^\prime\right)\ .
\end{eqnarray}
After introducing complex variables 
\begin{equation}
\label{vatra}
z_i = \exp\left(\frac{2\pi i x_i}{L}\right) \ ,\quad z^\prime_i = \exp\left(\frac{2\pi i x^\prime_i}{L}\right)
\end{equation}
Eq.~(\ref{recursiona}) becomes an integral representation for $\psi_{N+1}({\bf k}, {\bf z})$ 
\begin{eqnarray}
\label{jack1}
\psi_{N+1}({\bf k}, {\bf z})  &=&   \frac{C_{N}({\bf k})}{\sqrt{N+1}}
                                    \left(\frac{L}{2\pi i}\right)^N \int_{I^{(N)}_{\rm in}} d^N[{\bf z}^\prime]\nonumber\\
			& &     \left(\frac{\prod_{i=1}^{N+1}\prod_{j=i}^{N}(z_i-z^\prime_j)\prod_{j=1}^{N}\prod_{i=j+1}^{N+1}(z^\prime_j-z_i)}
			{\prod_{i<k}^{N+1}(z_i-z_k)\prod_{i<k}^{N}(z_i^\prime-z_k^\prime)}\right)^\lambda \nonumber\\
				&& (2i)^{-\lambda N}\prod_{i=1}^N {z_i^\prime}^{-\lambda-1-\frac{L k_{N+1}}{2\pi}}
				   \prod_{i=1}^{N+1} z_i^{+\frac{L k_{N+1}}{2\pi}}
				   \psi_{N}({\bf k},{\bf z}^\prime)  \ ,
\end{eqnarray}
with $d^N[{\bf z}^\prime]$ $=$ $\prod_{i=1}^N dz_i^\prime$. Using the form of Eq.~(\ref{jack1a}) 
for $\psi_{N}$ $({\bf k}, {\bf z}^\prime)$ as well as  
for $\psi_{N+1}$ $({\bf k}, {\bf z})$ on both sides of Eq.~(\ref{jack1}) 
\begin{eqnarray}
\label{jack2}
J^{(\lambda)}_{N+1}({\bf n}, {\bf z})  &=&  \frac{ C_{N}({\bf k})(2i)^{-\lambda N}}{\sqrt{N+1}}  
                             \left(\frac{L}{2\pi i}\right)^N 
                          \prod_{i=1}^{N+1} z_i^{\frac{Lk_{N+1}} {2\pi}+(\lambda+1)N} \nonumber\\
                        & &  \int_{I^{(N)}_{\rm in}} d^N[{\bf z}^\prime]\mu_\lambda({\bf z},{\bf z}^\prime)
			    \prod_{i=1}^N {z_i^\prime}^{-(\lambda+1)N-\frac{L k_{N+1}}{2\pi}}
				   J^{(\lambda)}_{N}({\bf n},{\bf z}^\prime) \ 
\end{eqnarray}
is obtained. We recall that $\mu_\lambda({\bf z},{\bf z}^\prime)$ was defined in Eq.~(\ref{anderson1}) in the introductory section. Here ${\bf n}$ is the same partition of length $N$ on both sides. Now periodicity of 
$a_{k_{N+1}}^\dagger({\bf x},{\bf x}^\prime)$ requires 
\begin{equation}
\label{nkrelation}
n_{N+1} = \frac{L}{2\pi}k_{N+1}+N(\lambda+1) \ ,\quad {\rm with} \ n_{N+1}\in {\mathbb Z} \ .
\end{equation}
This equation can be iterated
\begin{equation}
\label{nk1}
k_i \ =\ \frac{2\pi}{L}\left(n_i-(\lambda+1)(i-1)\right) \ ,
\end{equation}
where we assume that all $n_i \in {\mathbb N}$. We can restrict ourselves to positive integers, since any negative integers can be absorbed by an appropriate boost. 
Subtracting the center of mass momentum $K$ from every $k_i$ yields 
\begin{equation}
\label{bethe1a}
k_i \ =\ \frac{2\pi}{L}\left(n_i + \frac{\lambda+1}{2}(N+1-2i)\right) \ ,\quad  1\leq i\leq N+1\ . 
\end{equation}
It is easy to verify that the $k_i$ in Eq.~(\ref{bethe1a}) are solutions of the Bethe equation (\ref{bethte1}).
The integers $I_i$ are related to $n_i$ by $I_i = n_i + (N+1-2i)/2$. This completes the proof of Prop.~\ref{prop2} .

\begin{proposition}\label{prop3}
The action of the creation operator $\hat{a}_k^\dagger$ is equivalent to the integral representation for 
Jack polynomials, found by Olshanski and Okounkov in \cite{oko97} and in \cite{kuz03}. 
\end{proposition}

\paragraph{Proof}
In order to obtain the recursion formula for Jack polynomials as found in \cite{oko97}, we use Eq.~(\ref{nkrelation}) and the well known property of Jack polynomials
\begin{equation}
\left(\prod_{i=1}^Nz_i\right) J_N({\bf n},{\bf z}) = J^{(\lambda)}_N(\{n_1+1,n_2+1,\ldots\},{\bf z}) \equiv  J^{(\lambda)}_N({\bf n}+1,{\bf z}) \ .
\end{equation}
Eq.~(\ref{jack2}) is written as
\begin{eqnarray}
\label{jack3}
J^{(\lambda)}_{N+1}({\bf n}-n_{N+1}, {\bf z})  &=&  
                          \frac{ C_{N}({\bf k}) (2i)^{-\lambda N}\left(\frac{L}{2\pi i}\right)^N }
			  {\sqrt{N+1}} \nonumber\\
            &&\quad\int_{I^{(N)}_{\rm in}} d^N[{\bf z}^\prime]
			           \mu_\lambda({\bf z},{\bf z}^\prime)
				   J^{(\lambda)}_{N}({\bf n}-n_{N+1},{\bf z}^\prime) \ .
\end{eqnarray}
This is exactly the result by Okounkov and Olshanski (\cite{oko97}, Proposition 6)  for a Jack polynomial with partition ${\bf n}-n_{N+1}$, or equivalently for a Jack polynomial $J_{N,K-n_{N+1}}$ (boosted by $-n_{N+1}$) with partition ${\bf n}$, if we adjust the normalisation constants
\begin{equation}
C_{N}({\bf k}) (2i)^{-\lambda N}\left(\frac{L}{2\pi i}\right)^N \ =\ \prod_{i=1}^{N} [B(n_i-n_{N+1}+(N+1-i)(\lambda+1),\lambda+1)]^{-1}\ .
\end{equation}
Here $B(x,y)$ is Eulers beta--function. The quasimomenta $k_i$ on the l.~h.~s. are related to the integers $n_i$ on the r.~h.~s. by Eq.~(\ref{nk1}). This completes the proof of Prop.~\ref{prop3}.

\begin{proposition}\label{prop4}
Let $J^{(\lambda)}_{N}$ $({\bf n},{\bf z})$ be a Jack polynomial to partition ${\bf n} $ $=$ $\{n_1,\ldots,n_N\}$ and let 
$n$ be a positive integer. Define
\begin{equation}\label{vu}
\nu_{n,\lambda}({\bf z},{\bf z}^\prime) \ =\ \frac{\Delta_{N}({\bf z}^\prime)} {\Delta^{2\lambda+1}_{N-1}({\bf z})}
         \frac{\prod_{i=1}^{N-1} z^{-2\lambda-1}}{\prod_{i=1}^N {z_i^\prime}^{n+1}}
 				   \prod_{i=1}^{N-1}\prod_{j=i+1}^{N}(z_i-z^\prime_j)^\lambda\prod_{j=1}^{N}
                                    \prod_{i=j}^{N}(z^\prime_j-z_i)^\lambda \ .                            
\end{equation}
Then the following integral representation for Jack polynomials holds 
 \begin{equation}
 \label{annihil}
\oint dz_1^\prime \int_1^{z_1} dz_2^\prime\ldots \int_1^{z_{N-1}}dz_N^\prime 
                     \nu_{n,\lambda}({\bf z},{\bf z}^\prime)
                       J^{(\lambda)}_{N}({\bf n},{\bf z}^\prime) = B_N \delta_{n, n_m} J^{(\lambda)}_{N-1}({\bf n}-n,{\bf z})  \ .
\end{equation}
For $\lambda \in {\mathbb N}$, $n_m$ $=$ $n_i+ (\lambda+1)(N-i)$, $1\leq i\leq N$. 
For $\lambda \notin {\mathbb N}$, $m=N$.
The contour integral is over the unit circle.
The normalisation constant is 
\begin{equation}
B_N \ =\ 2\pi i
 \prod_{\genfrac{}{}{0pt}{}{i=1}{i\neq m}}^{N} [B(n_i-n_m+(m-i)(\lambda+1),\lambda+1)]\ .
\end{equation}
\end{proposition}

\paragraph{Proof}
The action of the annihilation operator $\hat{a}_k$ is defined by Eq.~(\ref{recursionb}). It is given in coordinate free notation by
\begin{equation}
 \hat{a}_k|\psi_N({\bf k})\rangle \ =\ \sum_{i} (-1)^{i+1}\delta_{k, k_i}|\psi_{N-1}({\bf k}_{\neq i})\rangle \ , 
\end{equation}
where in the state $|\psi_{N-1}({\bf k}_{\neq i})\rangle$ a particle with quasimomentum $k_i$ has been deleted. After the variable transformation (\ref{vatra}) using Eq.~(\ref{jack1a}) yields  
\begin{eqnarray}
\label{prop4a}
 J^{(\lambda)}_{N-1}({\bf n}, {\bf z})  &=& \frac{\sqrt{N}}{L} C_{N-1}({\bf k})(2i)^{-\lambda (N-1)}  
                             \left(\frac{L}{2\pi i}\right)^N 
                           \nonumber\\
                        & & \qquad\qquad \int_{I^{(N)}_{\rm out}} d^N[{\bf z}^\prime]
			           \nu_{n,\lambda}({\bf z},{\bf z}^\prime)
				   J^{(\lambda)}_{N}({\bf n},{\bf z}^\prime) \ ,
\end{eqnarray}
where we have set
\begin{equation}
n = \frac{L}{2\pi} k + (\lambda+1)(N-1)  \ .
\end{equation}
The relation of the set of integers ${\bf n}$ to the set of quasimomenta ${\bf k}$ is given by Eq.~(\ref{nk1}).  
Periodicity of $a_{k}({\bf x},{\bf x}^\prime)$ requires $n \in {\mathbb N}$. 
Therefore the Kronecker--delta for the quasimomenta transforms to 
\begin{equation}
\delta_{k, k_i} \ \to \ \delta_{n, n_i+(\lambda+1)(N-i)} \ .
\end{equation}
For $i=N$ this is just $\delta_{n, n_N}$. For $i\neq N$ it can only be non--zero for $\lambda\in {\mathbb N}$. This is one assertion of Prop. \ref{prop4}. Due to the antisymmetry of the integrand in Eq.~(\ref{recursionb}), the lower bounds in ${I^{(N)}_{\rm out}}$  can be extended for all integration variables $x_i^\prime$ to zero, respectively for all $z_i^\prime$ in Eq.~(\ref{prop4a}) to one, without changing the integral. This yields Eq.~(\ref{annihil}) and completes the proof of Prop.~\ref{prop4}.    

\subsection{ Rational Calogero--Moser--Sutherland system}
\label{rat}
For the type (II)--interaction potential the model is called rational CMS--model. Recursion formula (\ref{recursiona}) was derived 
in a slightly different form in \cite{guh02a}. There, instead of $\psi_N({\bf k},{\bf x})$, 
\begin{equation}
\phi_N({\bf k},{\bf x}) \ =\ \prod_{n<m}^N \sgn(k_n-k_m)\sgn(x_n-x_m)
\frac{\psi_N({\bf k},{\bf x})}{|\Delta_N({\bf x})\Delta_N({\bf k})|^{\lambda+1}}
\end{equation}
was considered. $\phi_N({\bf k},{\bf x})$ is completely symmetric in ${\bf x}$ 
and in ${\bf k}$, as well as under interchange 
of the two sets ${\bf x}$ and ${\bf k}$. For the special values $\lambda =$ $-1/2,0,1$ it is the group integral, see 
Eq.~(\ref{recurs}) of the introduction, 
\begin{equation} 
\phi^{(\lambda)}_N({\bf k},{\bf x})\ = \ \int_{U\in G(N)} d\mu(U)\exp(-i{\rm tr}U^{-1}{\bf x}U{\bf k}) \ .
\end{equation}
For $\lambda =-1/2$ the integration manifold $G(N)$ is the unitary group over the real field or equivalently the orthogonal group $O(N)$. For $\lambda = 0$, $G(N)$ is the unitary group over the complex field and for $\lambda = 1$, $G(N)$ is the unitary group over the quaternion field or equivalently the unitary symplectic group $USp(2N)$. The parameter $\beta$ of Ref.~\cite{guh02a} is related to the coupling constant $\lambda$ by $\lambda$ $=$ $\beta/2-1$ (see footnote \ref{fn1}).
The measure function 
\begin{equation}
 \mu_\lambda({\bf x},{\bf x}^\prime) \ = \ |\Delta_{N+1}({\bf x})|^{-\lambda-1} 
                     a^\dagger_0({\bf x},{\bf x}^\prime)|\Delta_{N}({\bf x}^\prime)|^{\lambda+1} 
\end{equation}
has the geometrical interpretation as the invariant 
Haar measure over the coset
\begin{equation}
\frac{\widehat{G}(N)}{\widehat{G}(N-1)}\ ,\quad \widehat{G}(N) \ = \ 
    \frac{G(N)}{\underbrace{G(1)\times\ldots \times G(1)}_{\rm N\ times}} \ ,
\end{equation}
in a special parametrization, called Gelfand--Tzetlin coordinates \cite{gel50,guh02a}. 
$G_1 \times G_2$ denotes the direct product group. Up to now there exist explicit results for wave functions of the rational CMS--model only for $\lambda=1$ for small particle number up to $N=4$  and for three particles for arbitrary $\lambda$ \cite{hik03}. Recently Bergere and Eynard derived a recursion formula similar to \eqref{recurs}, where the integration domains lie not on the real axis but are contour integrals \cite{ber09}. Using Cauchy's integral theorem they were able to derive more explicit forms for $\phi^{(\lambda)}_N({\bf k},{\bf x})$ in the cases of $\lambda$ integer. In particular for the value $\lambda=1$ corresponding to the group integral over the unitary symplectic group a rather explicit expression was obtained. 

\subsection{Hyperbolic CMS system}
\label{hyp}

The hyperbolic CMS Hamiltonian, model (III), has been investigated by Sutherland with the asymptotic Bethe Ansatz. By now a rather profound understanding of the physics of the system has been obtained \cite{sut04}. However, in contrast to the trigonomatric CMS model, relatively little is known about its eigenfunctions. It is beyond the scope of this work to fully work out the eigenfunctions to the hyperbolic CMS Hamiltonian. Future research will show, whether or not this goal can be achieved using the recursive formalism presented here.    

However, it is instructive to see, how the two--particle wave function can be obtained by acting with the creation operator $a^\dagger_{k_2}$ onto a one particle state, i.~e. onto a plane wave. 
Introducing new variables 
$z_i$ $=$ $\exp\left(2a x_i\right)$, $i=1, 2$ and $z^\prime_i$ $=$ $\exp\left(2a x^\prime_i\right)$
the symmetric creation function $a^\dagger_{k_2}({\bf z},z^\prime)$ reads 
\begin{eqnarray}
a^\dagger_{k_2}({\bf z},z^\prime)dz^\prime &=& e^{ik_2(x_1+x_2)} (z^\prime)^{-ik_2/2a -\lambda-1}
                       \frac{|z_1-z^\prime|^\lambda|z^\prime-z_2|^\lambda}{(2|z_1-z_2|)^\lambda}\frac{dz^\prime}{2a}
\end{eqnarray}
Applying recursion formula \eqref{recursiona} one finds for the two--particle wave function  
\begin{eqnarray}
\label{psi2a}
\psi^{\rm (III)}_2({\bf k},{\bf z})  &=& \frac{C_1({\bf k})}{\sqrt{2}}\frac{e^{i(k_1+k_2)x_2}}{2a}
                                  \frac{ {\cal F}^{\rm (III)}(z_1,z_2) - {\cal F}^{\rm (III)}(z_2,z_1)}
                                  {(2|z_1-z_2|)^\lambda} \ ,                      
\end{eqnarray}
where 
\begin{eqnarray}
\label{psi2b}
{\cal F}^{\rm (III)}(z_1,z_2) & = & \int^{z_1}_{0}dz^\prime (z^\prime)^{ik^\prime-\lambda-1}
                       (z_1-z^\prime)^\lambda(z^\prime-z_2)^\lambda\\
                   &=& B(ik^\prime-\lambda,\lambda+1) z_1^{ik^\prime}(-z_2)^\lambda 
                       F(-\lambda,ik^\prime-\lambda,ik^\prime+1;z_1/z_2).   
\end{eqnarray}
Here we recall that $B(x,y)$ is Euler's beta function and $F(a,b,c;z)$ is the hypergeometric function as defined in \cite{abr72}. Moreover we defined $k^\prime$ $=$ $(k_1-k_2)/(2a)$. The expressions \eqref{psi2a} and \eqref{psi2b} can be evaluated further in the asymptotic regime $z_1\gg z_2$. We use the asymptotic formula for the hypergeometric function
\begin{equation}
\label{Fasymp}
F(a,b,c;z) \stackrel{z\to\infty}{\longrightarrow}\frac{\Gamma(c)\Gamma(b-a)}{\Gamma(b)\Gamma(c-a)}(-z)^{-a}+ \frac{\Gamma(c)\Gamma(a-b)}{\Gamma(a)\Gamma(c-b)}(-z)^{-b} \ .
\end{equation}
The two--particle wave function becomes in this limit
\begin{eqnarray}
\label{psi2c}
\psi_2^{\rm (III)}({\bf k},{\bf z})  & \stackrel{z_1\gg z_2}{\longrightarrow}& \frac{C_1({\bf k})}{\sqrt{2}}\frac{B(ik^\prime,\lambda+1)}{2^{\lambda+1} a}\nonumber\\
               &&\quad \left(e^{ik_1x_1+ik_2x_2} + S^{\rm (III)}(k^\prime) e^{ik_1x_2+ik_2x_1}\right) \ .  
\end{eqnarray}
where the two--body $S$--matrix $S(k^\prime)$ is found to be 
\begin{eqnarray}
\label{psi2d}
S^{\rm (III)}(k^\prime) &=& \frac{B(i k^\prime-\lambda,-i k^\prime)}{B(i k^\prime,\lambda+1)}(-1)^{-i\pi k^\prime} - 
                \frac{B(i k^\prime-\lambda,\lambda+1)}{B(i k^\prime,\lambda+1)}(-1)^{-\lambda}\nonumber\\
             &=& - \frac{\Gamma(1-i k^\prime)\Gamma(\lambda+1+i k^\prime)}
                       {\Gamma(1+i k^\prime)\Gamma(\lambda+1-i k^\prime)}\ .
\end{eqnarray}
The second equation was obtained using the reflection formula for the Gamma function $\Gamma(x)\Gamma(1-x)$ $=$ $\pi/\sin(\pi x)$. The limit $a\to \infty$ corresponds to an infinitely small interacting region. In the limit $a\to \infty$, $\lambda\to 0$ and $ a\lambda = c$ the $S$--matrix is $S^{\rm (III)}(k_1-k_2) = - 1$. The inverse--$\sinh$ interaction becomes invisible for Fermions.

For an interaction potential of the Morse type, model (IV), the wave function 
\begin{eqnarray}
\label{psi3a}
\psi^{\rm(IV)}_2({\bf k},{\bf z})  &=& \frac{C_1({\bf k})}{\sqrt{2}}\frac{e^{i(k_1+k_2)x_2}}{2a}
                                  \frac{ {\cal F}^{\rm(IV)}(z_1,z_2) - {\cal F}^{\rm(IV)}(z_2,z_1)}
                                  {(2|z_1+z_2|)^\lambda} \ ,                      
\end{eqnarray}
is obtained, where
\begin{eqnarray}
\label{psi3b}
{\cal F}^{\rm (IV)}(z_1,z_2) & = & \int^{z_1}_{0}dz^\prime (z^\prime)^{ik^\prime-\lambda-1}
                       (z_1+z^\prime)^\lambda(z^\prime+z_2)^\lambda \ . 
\end{eqnarray}
The integral can be expressed by a yet more general hypergeometric function of two arguments \cite{gra80}. Using an asymptotic expansion similar to \eqref{Fasymp} we find in the region $z_1\gg z_2$
\begin{eqnarray}
\label{psi3c}
\psi^{\rm(IV)}_2({\bf k},{\bf z})  & \stackrel{z_1\gg z_2}{\longrightarrow}& \frac{C_1({\bf k})}{\sqrt{2}}
   \frac{F(-\lambda,ik^{\prime},ik^\prime+1;-1)}{2^{\lambda}i(k_1-k_2)}\nonumber\\
               &&\quad \left(e^{ik_1x_1+ik_2x_2} + S^{\rm (IV)}(k^\prime) e^{ik_1x_2+ik_2x_1}\right)\ ,                      
\end{eqnarray} 
where the $S$--matrix is now given by the more complicated expression
\begin{eqnarray}
\label{Scosh}
S^{\rm (IV)}(k^\prime) &=& S^{\rm (III)}(k^\prime)\frac{\sin(\pi[\lambda+1])+ \sin(i\pi k^\prime)}{\sin(\pi[\lambda+1-ik^\prime])} \ .
\end{eqnarray}
Properties of this scattering matrix are discussed in detail elsewhere \cite{sut04}. For $\lambda = 0 $ the extra term in \eqref{Scosh} becomes one and $S^{\rm (IV)}=S^{\rm (III)}$. In the limit of $\delta$--interaction the two--body $S$--matrix becomes again $S^{\rm (IV)}(k^\prime) = -1$ as in the hyperbolic CMS model, however in a more complicated way.

\section{Proofs}
\label{proof}
We prove Theorem \ref{theo0} and Proposition \ref{prop1}
\subsection{Proof of Theorem \ref{theo0} }
\label{proof11}
We prove the five points of Theorem \ref{theo0}

1.) In order to prove parts \ref{item1} and \ref{item2} of Theorem \ref{theo0} we need to 
show that the symmetric annihilation function 
$a_k({\bf x},{\bf x}^\prime)$ and the creation function
 $a^\dagger_k({\bf x},{\bf x}^\prime)$ defined in (\ref{theo3}) are solutions of the 
differential equations (\ref{res6coord}).
 We first focus on the creation function $a^\dagger_k({\bf x},{\bf x}^\prime)$.
 We make for $a^\dagger_k({\bf x},{\bf x}^\prime)$ the Ansatz 
 \begin{eqnarray}
 \label{proof1}
 a_k^\dagger({\bf x},{\bf x}^\prime) &=& \exp\left[-\sum_{n< m}^{N+1} F(x_n-x_m)+\sum_{n, m} F(x_n-x^\prime_m)-\right.\nonumber\\
&&\ \left. \sum_{n< m}^N F(x^\prime_n-x^\prime_m)+ ik\left(\sum_{n=1}^{N+1} x_n-\sum_{m=1}^{N}x_m^\prime\right)\right] \ ,
 \end{eqnarray}
where $F(x)$ is an arbitrary even function. Acting with $\sum_{n=1}^{N+1}\frac{\partial^2}{\partial x_n^2}$ and with $\sum_{n=1}^{N}\frac{\partial^2}{\partial {x_n^\prime}^2}$ on $a_k^\dagger({\bf x},{\bf x}^\prime)$ yields
 \begin{eqnarray} \label{proof2}   
&&\sum_{n=1}^{N+1}\frac{\partial^2}{\partial x_n^2} a_k^\dagger({\bf x},{\bf x}^\prime) \ =\
    \left\{\sum_{n\neq m}\left([f^2(x_n-x_m)-f^{\prime}(x_n-x_m)\right)+\right.\nonumber\\ 
    &&\quad\sum_{l\neq m\neq n}^N f(x_n-x_m) f(x_n-x_l)-
    2\sum_{n\neq m}^{N+1}\sum_l^{N} f(x_n-x_m) f(x_n-x^\prime_l)+\nonumber\\
    &&\quad\sum_{l\neq m}^{N}\sum_n^{N+1} f(x_n-x^\prime_m) f(x_n-x^\prime_l)- (N+1)k^2+ \\ 
    &&\left. \sum_n^{N+1}\sum_{m}^{N} \left[f^2(x_n-x^\prime_m)+ f^\prime(x_n-x^\prime_m) +
    ik f(x_n-x^\prime_m)\right] \right\} a_k^\dagger({\bf x},{\bf x}^\prime)\ ,\nonumber
\end{eqnarray}
and, by the same token
 \begin{eqnarray}
  \label{proof3}  
&&\sum_{n=1}^{N}\frac{\partial^2}{\partial {x^\prime_n}^2} a_k^\dagger({\bf x},{\bf x}^\prime) \ =\
    \left\{\sum_{n\neq m}\left(f^2(x^\prime_n-x^\prime_m)- f^{\prime}(x^\prime_n-x^\prime_m)\right)+\right.\nonumber\\ 
    &&\quad\sum_{l\neq m\neq n}^N f(x^\prime_n-x^\prime_m) f(x^\prime_n-x^\prime_l)+
    2\sum_{l\neq m}^{N}\sum_n^{N+1} f(x^\prime_l-x^\prime_m) f(x_n-x^\prime_l)+\nonumber\\
    &&\quad\sum_{n\neq m}^{N+1}\sum_l^{N} f(x_n-x^\prime_l)f(x_m-x^\prime_l)- Nk^2+\nonumber\\
    && \left. \sum_n^{N+1}\sum_{m}^{N} \left[f^2(x_n-x^\prime_m)+
    f^{\prime}(x_n-x^{\prime}_m)+ ik f(x_n-x^\prime_m) \right] \right\}
    a_k^\dagger({\bf x},{\bf x}^\prime)\ ,  
\end{eqnarray}
where we defined $f(x)=\frac{d}{dx}F(x)$ and  $f^\prime(x)=\frac{d^2}{dx^2}F(x)$. 
We now define the Hamiltonian $\widetilde{H}_{N}$ in ${\bf x}$ representation, as
\begin{eqnarray} \label{proof4}
\widetilde{H}_{N}({\bf x}) &=& -\sum_{n=1}^{N}\frac{\partial^2}{\partial {x_n}^2}+
                \sum_{n\neq m}\left(f^2(x_n-x_m)-f^{\prime}(x_n-x_m)\right)+\nonumber\\ 
                  &&\sum_{l\neq m\neq n}^N f(x_n-x_m)f(x_n-x_l)\ ,
\end{eqnarray}
which comprises all terms in Eqs.~(\ref{proof2}) and (\ref{proof3}), which depend only on one set of variables.
Subtracting Eq.~(\ref{proof2}) from (\ref{proof3}) yields
 \begin{eqnarray} \label{proof5}
&&\left[\widetilde{H}_{N+1}({\bf x}) - \widetilde{H}_{N}({\bf x}^\prime)\right] a_k^\dagger({\bf x},{\bf x}^\prime) \ =\  k^2 a_k^\dagger({\bf x},{\bf x}^\prime) + \nonumber \\ 
&&  \left\{\sum_{l\neq m}^{N}\sum_n^{N+1} 
\left[ 2 f(x^\prime_l-x^\prime_m) f(x_n-x^\prime_l)-f(x_n-x^\prime_m) f(x_n-x^\prime_l)\right] + \right. \\
&&\left.\sum_{n\neq m}^{N+1}\sum_l^{N}\left[ f(x_n-x^\prime_l)f(x_m-x^\prime_l)+2f(x_n-x_m) f(x_n-x^\prime_l)\right]\right\}
a_k^\dagger({\bf x},{\bf x}^\prime).\nonumber
\end{eqnarray}
One now might conclude that the functional equation
\begin{eqnarray}\label{appc1}
&&f(x_1-x_2)f(x_1-y)+ f(x_2-x_1)f(x_2-y) +f(x_1-y)f(x_2-y) \ = \nonumber\\
&&\qquad\qquad \qquad  v(x_1,x_2)+u(y)
\end{eqnarray}
with arbitrary functions $v$ and $u$ is a sufficient condition on $f$. In this case $v(x_1,x_2)$ would be an additional two--body interaction term and $u(y)$ an additional single particle term in the potential. However this functional equation can be simplified further by invoking translation invariance. Translation invariance of the left hand side of \eqref{appc1} requires
translation invariance of the right hand side
\begin{eqnarray}
\label{appc2}
v(x_1+a,x_2+a) + u(y+a) &=& v(x_1,x_2) + u(y)\nonumber\\
\left(\frac{d}{dx_1}+\frac{d}{dx_2}\right)v(x_1,x_2) + \frac{d}{dy} u(y) &=& 0 \ .
\end{eqnarray}
The second equation has two solutions. Either $v(x_1,x_2)=v(x_1-x_2)$ and $u(y)={\rm const.}$, or $v(x_1,x_2) = c_1(x_1+x_2)$ and $u(y) = -2 c_1 y$ with an arbitrary constant $c_1\in {\mathbb R}$. We first focus on the first solution. It means that the right hand side of Eq.~\eqref{appc1} is independent of $y$. Now one can invoke 
permutation symmetry of the left hand side in Eq.~(\ref{appc1}) under interchanging $x_1\leftrightarrow y$ and
 $x_2 \leftrightarrow y$. The same symmetry must hold on the right hand side, and therefore
 $v(x_1,x_2)={\rm const.}$, too. Eq.~\eqref{appc1} reduces to  
\begin{eqnarray}
f(x)f(y)+f(x)f(z)+f(y)f(z) & = & {\rm const.}\nonumber\\
x+y+z & = &0  , \label{appc1a}
\end{eqnarray} 
which is the functional equation \eqref{theo1} of theorem \ref{theo0}.

If $f(x)$ fulfills \eqref{appc1a}
the term in the squared bracket on the left hand side of \eqref{proof5} becomes constant. 
We can rewrite Eq.~\eqref{proof5} as
 \begin{eqnarray} \label{proof6}
\left[\widetilde{H}_{N+1}({\bf x}) - \widetilde{H}_{N}({\bf x}^\prime)\right] a_k^\dagger({\bf x},{\bf x}^\prime) & = & 
       \left[ k^2-N(N+1) {\rm const.}\right] a_k^\dagger({\bf x},{\bf x}^\prime) \ .
\end{eqnarray}
Observing that condition \eqref{appc1a} also yields
\begin{equation}
\label{threeparticle}
 \sum_{l\neq m\neq n}^N f(x_n-x_m)f(x_n-x_l) \ =\ - \frac{1}{3}N(N-1)(N-2){\rm const.}\ ,
\end{equation}
we find 
\begin{eqnarray}
k^2  a_k^\dagger({\bf x},{\bf x}^\prime) & = & 
\left[H_{N+1}({\bf x}) - H_{N}({\bf x}^\prime)\right] a_k^\dagger({\bf x},{\bf x}^\prime)\\
H_{N+1}({\bf x}) &=&   -\sum_{n=1}^{N+1}\frac{\partial^2}{\partial x_n^2} + \sum_{n\neq m} V(x_n-x_m)\nonumber\\ 
 V(x_n-x_m)      &=&     f^2(x_n-x_m)-f^\prime(x_n-x_m)+{\rm const.}.
\end{eqnarray}
This is our assertion Eq.~\eqref{potential} with dispersion relation
\begin{equation}\label{dis}
\epsilon(k)\ =\ k^2 \ .
\end{equation}
Performing the same analysis with the second solution of \eqref{appc2}, yields an awkward looking $N$ dependent single particle term in the potential, which seems not to describe realistic situations. We therefore discard it.
 
The proof for the annihilation function $a_k({\bf x},{\bf x}^\prime)$ goes along the same lines.
This completes the proof of part \ref{item1} and of part \ref{item2} of Theorem \ref{theo0}.

2.) In order to prove the recursion formula in part \ref{item3} we have to show that  $H_{N+1}$ ($H_{N-1}$) commute with the statistical functions $I_N^\dagger({\bf x},{\bf x}^\prime)$ ($I_N({\bf x},{\bf x}^\prime)$), such that 
\begin{eqnarray}
\label{recursionc1}
H_{N+1}({\bf x}) I_N^\dagger ({\bf x},{\bf x}^\prime) a_{k}^\dagger({\bf x},{\bf x}^\prime) &=&  I_N^\dagger ({\bf x},{\bf x}^\prime) a_{k}^\dagger({\bf x},{\bf x}^\prime) H_{N}({\bf x}^\prime)\nonumber\\
H_{N-1}({\bf x}) I_N({\bf x},{\bf x}^\prime) a_{k}({\bf x},{\bf x}^\prime) &=&  I_N({\bf x},{\bf x}^\prime) a_{k}({\bf x},{\bf x}^\prime) H_{N}({\bf x}^\prime)\ .
\end{eqnarray}
We focus on the first equation. We act with $H_{N+1}({\bf x})$ on $I_N^\dagger ({\bf x},{\bf x}^\prime) 
a_{k}^\dagger({\bf x},{\bf x}^\prime)$ use product rule of differentiation, part \ref{item1} of theorem \ref{theo0} and integrate by parts. We find that Eq.~\eqref{recursionc1} is fulfilled, if $I_N^\dagger({\bf x},{\bf x}^\prime)$ meets the following condition
\begin{eqnarray}
\label{Iprop0}
\sum_{n=1}^{N+1}\left(2\frac{\partial I_N^\dagger}{\partial x_n}\frac{\partial a_k^\dagger}{\partial x_n}+
a_k^\dagger\frac{\partial^2 I_N^\dagger}{\partial x_n^2}\right) &=& - \sum_{m=1}^{N}
  \left( a_k^\dagger\frac{\partial I_N^\dagger}{\partial x_m^{\prime 2}}+
2 a_k^\dagger\frac{\partial I_N^\dagger}{\partial x_m^\prime}\frac{\partial}{\partial x_m^\prime}\right)\ .          
\end{eqnarray}
Using the explicit form \eqref{stat1} of $I_N^\dagger({\bf x},{\bf x}^\prime)$, we see that derivatives of $I_N^\dagger$ with respect to an unprimed argument can be transformed into derivatives with respect to a primed argument
\begin{eqnarray}
\label{Iprop1}
\frac{\partial}{\partial x_n} I_N^\dagger &=& \frac{2^{-(N+1)}}{(N+1)!}\det\left|\begin{matrix}
                \D \ldots& \D \ldots& \D \ldots& \D \ldots\cr
                  \D \sgn(x_n-x^\prime_1)\frac{\partial}{\partial x_1^\prime}
            &\ldots&\D\sgn(x_n-x^\prime_N)\frac{\partial}{\partial x_N^\prime}&\D 1\cr
				 \D \ldots &\D \ldots & \D \ldots& \D \ldots
                                    \end{matrix}\right| 
\end{eqnarray}
where all other entries of the determinant remain unchanged. Using Eq.~\eqref{Iprop1} and 
\begin{equation}
\delta(x_j^\prime-x_i)\frac{\partial}{\partial x_{i}}a^\dagger_{k}({\bf x}^\prime,{\bf x}) \ =\ - \delta(x_j^\prime-x_i)\frac{\partial}{\partial x_{j}^\prime}a^\dagger_{k}({\bf x}^\prime,{\bf x})
\end{equation}
it is straightforward to show that Eq.~\eqref{Iprop0} holds. The calculation for the annihilation operator goes 
along the same lines. 

We next show that $a_k$ is an annihilation operator, i.~e.~
\begin{equation}
\label{annihilcond}
a_k|\psi_N({\bf k})\rangle \ =\ 0 \ ,\quad \mbox{if $k\neq k_1\ldots k_N$} \ .
\end{equation}
To this end (for the moment) we assume part \ref{item5} of Theorem \ref{theo0} to be proven and 
consider the pairing
\begin{eqnarray}
\label{scalar}
\langle \psi_N({\bf k^\prime}) |\psi_N({\bf k})\rangle &=& \left\{\begin{matrix}
                                         \displaystyle{\det [2\pi \delta(k^\prime_i-k_j)]_{1\leq i,j\leq N}}\cr
					 \displaystyle{\det [L \delta_{k^\prime_i, k_j}]_{1\leq i,j\leq N}}
                                                            \end{matrix}\right.\nonumber\\
              &= &\int_{\Omega^N} d [{\bf x}] \psi_N^*({\bf k}^\prime,{\bf x})\psi_N({\bf k},{\bf x}) \nonumber\\
 &=&\int_{\Omega^N}  d [{\bf x}]\int_{I^{(N)}_{\rm in}} 
         d [{\bf x}^\prime] \int_{I^{(N)}_{\rm in}} d [{\bf x}^{\prime\prime}]  
      \left[a^\dagger_{k^\prime_N}({\bf x},{\bf x}^\prime)\right]^*\psi_{N-1}^*({\bf k}^\prime,
  {\bf x}^{\prime})\nonumber\\
   & &\qquad a^\dagger_{k_N}({\bf x},{\bf x}^{\prime\prime}) \psi_{N-1}({\bf k},{\bf x}^{\prime\prime})\nonumber\\
    &=& \int_{\Omega^N}  d [{\bf x}^\prime ]
     \psi_{N-1}^*({\bf k}^\prime,{\bf x}^{\prime}) \int_{I^{(N)}_{\rm out}} d [{\bf x}]  
           a_{k^\prime_N}({\bf x}^\prime,{\bf x}) 
     \int_{I^{(N)}_{\rm in}} d [{\bf x}^{\prime\prime}] \nonumber\\
      & &\qquad a^\dagger_{k_N}({\bf x},{\bf x}^{\prime\prime}) 
        \psi_{N-1}({\bf k},{\bf x}^{\prime\prime})\nonumber\\
    & = &  \langle \psi_{N-1}({\bf k}^\prime) |  a_{k^\prime_N}  \psi_{N}({\bf k})\rangle  \nonumber\\
    &=& 0 \quad ,\quad {\rm if} \ k_N^\prime\neq k_1\ldots k_N \ .
\end{eqnarray}
This proves Eq.~(\ref{annihilcond}), since Eq.~(\ref{scalar}) holds for an arbitrary wave function 
$|\psi_{N-1}({\bf k}^\prime)\rangle$. 

3.)The pieces of part \ref{item4} of Theorem \ref{theo0} which concern the action of the Hamiltonian have been proven already before. To complete the proof we have to show in addition that a differential equation similar to Eq.~(\ref{res6coord}) holds for the center of mass momentum operator $P_{N}({\bf x})$, defined in Eq.~(\ref{com}), namely
\begin{eqnarray}
\left[P_{N+1}({\bf x}) - P_{N}({\bf x}^\prime)\right] a_k^{\dagger}({\bf x},{\bf x}^\prime)  
                  &= & k a_k^{\dagger}({\bf x},{\bf x}^\prime)\nonumber\\
\left[P_{N}({\bf x}) - P_{N+1}({\bf x}^\prime)\right] a_k({\bf x},{\bf x}^\prime)  
                  &= & - k a_k({\bf x},{\bf x}^\prime)	\ .
\end{eqnarray}
It is straightforward to see that this is true. 

4.) Finally we have to prove the orthogonality relation (i.~e.~part \ref{item5} of Theorem \ref{theo0}). 
To this end we observe that the $N$ real numbers 
${\bf k}$ $=$ $\{k_1,\ldots,k_N\}$ are conserved quantities. This means that exactly $N$ mutually commuting selfadjoint operators 
$I_n$ , $1\leq n\leq N$ can be constructed with eigenvalues $E_n$ $=$ $\sum_{i=1}^N k_i^n$. For the 
potentials in Table \ref{table1} these operators can be constructed with Dunkl operators $p_m$
\cite{dun89,pol92}
\begin{eqnarray}
I_n &=& \sum_{m=1}^N p_m^n\nonumber\\
p_m &=& -i\frac{\partial}{\partial x_m} + \sum_{n\neq m} f(x_m-x_n)P_{nm} \ ,
\end{eqnarray} 
where $P_{nm}$ is the exchange operator defined in Eq.~(\ref{exchange}).
Therefore, for any two unequal sets ${\bf k}$ and ${\bf k}^\prime$ at least one selfadjoint 
operator $I_n$ can be found for which 
\begin{equation}
I_n \psi_N({\bf k},{\bf x}) \neq  I_n \psi_N({\bf k}^\prime,{\bf x}) \ , \quad {\rm if} \ {\bf k}\neq {\bf k}^\prime \ .
\end{equation}
Now we can invoke a fundamental theorem of functional analysis for selfadjoint operators: Two eigenfunctions of a 
selfadjoint operator to different eigenvalues are orthogonal. This completes the proof.

\subsection{Proof of Proposition \ref{prop1}}
\label{AppB}
The five potentials of Table \ref{table1} have to be treated differently. 

\paragraph{Potential (I) and (II)}  

For the trigonometric potential (I) the normalisation constant 
\begin{equation}
\label{potential1result}
C_N({\bf k})\ =\ \sqrt{N+1}\left(\frac{L}{2\pi i}\right)^{-N}
                        \prod_{i=1}^{N} \frac{(2i)^\lambda}{B\left(\frac{L}{2\pi}(k_i-k_{N+1}),\lambda+1\right)} \ .
\end{equation}
has been calculated already in Sec.~\ref{trig}. To obtain the normalisation constant for the rational potential (II) one can either follow the route of Sec.~\ref{propproof1} and perform the integration in the asymptotic regime. 
But the evaluation of the resulting integral is by no means trivial. Therefore we resort to a different route.
 
Eigenfunctions for the rational CMS Hamiltonian (II) emerge from the trigo\-no\-metric case (I) in the limit 
$L\to \infty$ keeping ${\bf k}$ and ${\bf x}$ finite. The same happens to their creation functions
\begin{equation}
a_k^{\rm (II)}({\bf x},{\bf x}^\prime) \ =\ \lim_{L\to\infty}\left(\frac{L}{\pi} \right)^{N\lambda}
                                                   a_k^{\rm (I)}({\bf x},{\bf x}^\prime) \ ,
\end{equation}
where the upper index denotes the type of potential.
This allows us to obtain the normalisation for the potential (II) case by taking the limit $L\to\infty$ of Eq.~(\ref{potential1result}) using the asymptotic expansion of the beta--function 
\begin{equation}
\lim_{x\to\infty} \frac{1}{B(x,y)} \ =\ \frac{x^y}{\Gamma(y)}(1+ {\rm lower \ order \ terms}) \ .
\end{equation}
The result is
\begin{equation}
C_N({\bf k}) \ =\  \frac{i^{N(\lambda+1)}}{\Gamma^N(\lambda+1)} 
                 \prod_{i=1}^N|k_i-k_{N+1}|^\lambda\left(k_i-k_{N+1}\right) \ .
\end{equation}
It differs from the normalisation Eq.~(5.7) of Ref.~\cite{guh02a} by a $k$ independent factor.

\paragraph{Potentials (III)--(V)} 
\label{propproof1}
For the $\delta$--interaction potential (V) the normalisation has already been calculated in Sec.~\ref{delta}. 
For the potentials (III) and (IV) the normalisation constant $C_N({\bf k})$ is determined by the condition that in the asymptotic regime, the wave function $\psi_{N}({\bf k},{\bf x})$ obtains the form of a scattering wave solution as given in Eq.~\eqref{asymp}. For integrable systems it is known that the $N$--body $S$--matrix is a product of two--body scattering matrices. In the following, however, we do not need this property. 

In order to determine the normalisation constant $C_N({\bf k})$ we look on the asymptotic behavior of the creation function. Introducing as in section \ref{hyp} variables $z_n$ $=$ $\exp\left(2a  x_n\right)$, $n=1, \ldots N\pm 1$ and $z^\prime_m$ $=$ $\exp\left(2a x^\prime_m\right)$, $m=1,\ldots N$ the creation function reads
\begin{eqnarray}
a_{k_{N+1}}({\bf z},{\bf z}^\prime)d[{\bf z}^\prime] &=& (2a)^{-N}\prod_{m=1}^N 2^{-\lambda}(z_m^\prime)^{-ik_{N+1}/(2a)-\lambda-1}
                                                    \prod_{n=1}^{N+1} z_n^{ik_{N+1}/(2a)}\nonumber\\
   && \frac{\prod_{n=1}^{N+1}\prod_{m=1}^N|z_n\pm z_m^\prime|^\lambda}{\prod_{n<m}^{N+1}|z_n\pm z_m|^\lambda
                                                    \prod_{n<m}^N|z_n^\prime \pm z_m^\prime|^\lambda}d[{\bf z}^\prime] \ .
\end{eqnarray}
Here the plus sign denotes interaction via the Morse potential, model (IV), and the minus sign denotes the hyperbolic CMS model (III). We first integrate over $z_1^\prime$ approximately in the limit $z_1\gg z_2\gg\ldots z_{N+1}$. Rescaling $z_1^\prime \to z_1 t$ yields
\begin{eqnarray}
\psi^{\rm (asym)}_{N+1}({\bf k},{\bf x})& = & \frac{C_N({\bf k})}{\sqrt{N+1}}
\frac{2^{-N\lambda}e^{ik_{N+1}\sum_{n=1}^{N+1}}}{(2a)^{N}}\int_{z_2/z_1}^1 dt t^{-ik_{N+1}/(2a)-\lambda-1}\nonumber\\
& & z_1^{-ik_{N+1}/(2a)}(1-t)^\lambda\frac{\prod_{n=2}^{N+1}(t\pm z_n/z_1)^\lambda}
                                     {\prod_{m=2}^{N}(t\pm z^\prime_m/z_1)^\lambda}
\psi_{N}({\bf k},{\bf z}^\prime)\ldots
\end{eqnarray}
where the dots denote the remaining integrals and terms in the creation function which do not depend on $t$. For $z_1\gg z_2$ we can extend the integration domain to the intervall $[0,1]$. Moreover we can substitute $\psi_{N}({\bf k},{\bf z}^\prime)$ by its asymptotic form \eqref{asymp}. Since we are interested only in the normalistation, it suffices to focus on a single term in the sum over the permutation group, say the unit element. We find
\begin{eqnarray}
\psi^{\rm (asym)}_{N+1}({\bf k},{\bf x})& = & \frac{C_N({\bf k})}{\sqrt{N+1}}
\frac{2^{-N\lambda}}{(2a)^{N}}e^{ik_1 x_1+ i k_{N+1}\sum_{n=2}^{N+1}}\nonumber\\
& & \int_{0}^1 dt t^{i(k_1-k_{N+1})/(2a)-1}(1\pm t)^\lambda\int(\ldots)\ ,
\end{eqnarray}
where the dots again denote all remaining terms. Now the $t$--integration can be performed using
\begin{equation}
\int_{0}^1 dt\, t^{ik-1}(1\pm t)^\lambda \ =\ B(ik,ik+1)F(-\lambda,ik,ik+1;\mp 1) \ .
\end{equation}    
Iterating the same procedure for the integration over $z_2^\prime\ldots z_N^\prime$ and adjusting $C_N({\bf})$ yields the result as stated in proposition \ref{prop1}. This completes the proof for potentials (III) and (IV).

\section{Conclusions}
\label{concl}

We constructed creation and an\-nihil\-ation oper\-ators for spinless interacting Fer\-mions. 
Applying the creation operators successively onto the vacuum any $N$--particle eigenstate can thereby be generated. The eigenstates are given as a $(N-1)N/2$--fold integral. For the trigonometric CMS--Hamiltonian the equivalence 
of these eigenfunctions to other representations of the eigenstates has been demonstrated.   

The developed formalism paves a new way of searching and classifying integrable quantum systems, complementary or alternative to Bethe's Ansatz and to the Yang--Baxter equation. An interacting many--body Hamiltonian is 
exactly solvable if it can be transformed by a unitary transformation 
to a Hamiltonian containing one--body operators only. The formalism might be useful for the calculation of correlation functions. The method is yet to be developed.

It has to be stressed that the constructed operators have always Fermionic commutation relations. Therefore they 
differ fundamentally from the Bosonic operators which appear for instance in the Bosonisation approach. The latter describes the fundamental excitations of the $N$ particle system by Bosonic operators which act on top of 
the filled Fermi sea. Thereby the Hamiltonian is effectively diagonalised. In contrast our diagonalisation is exact.

The constructed operators are similar but not equal to the Fadeev--\-Zamolod\-chikov operators \cite{kon04}. The latter obey commutation relations which involve the two--body scattering matrix and need not necessarily have the commutation relations (\ref{res6}) with the Hamiltonian. 

In the scattering approach based on the Fadeev--\-Zamolod\-chikov algebra the interaction becomes manifest in the commutation relations of the operators. In our approach the operators have Fermionic commutation relation. The interaction becomes manifest in the quantisation condition on the quasimomenta when periodic boundary conditions are introduced.  Our approach shares this feature with the original coordinate Bethe Ansatz \cite{bet31,yan67}.  
More on the connection of the creation (annihilation) operators, constructed here, to the Fadeev--Zamolodchikov algebra will be given elsewhere.

In the present work we focused on non--relativistic spinless Fermions. Moreover translation 
invariance was assumed. It has to be stressed that the developed formalism does not hinge on these assumptions. The extensions to spin $1/2$ Fermions and to Bosons will be given in separate publications. 
Application to lattice theories and 1--d relativistic field theories present an interesting challenge for the future.

\thanks
I acknowledge financial support by the German Research Council (DFG) with personal grants No. KO3538/1-1, KO3538/1-2 and within the SFB--TR/12. I thank F.~Calogero, B.~Gutkin, A.~Komnik, A.~Osterloh and C.~Recher for many helpful comments. I thank M.~Fury for proofreading the manuscript. 
 

\end{document}